\documentclass[a4paper]{amsart}

\RequirePackage{amsmath}
\RequirePackage{bm}
\RequirePackage{amssymb}
\RequirePackage{upref}
\RequirePackage{amsthm}
\RequirePackage{enumerate}
\RequirePackage{pb-diagram}
\RequirePackage{amsfonts}
\RequirePackage[mathscr]{eucal}
\RequirePackage{verbatim}
\RequirePackage{xr}
\RequirePackage{graphicx}
\usepackage{calc}
\usepackage{xspace}
\RequirePackage{color}
\RequirePackage{ifthen}
\RequirePackage{fancybox}
\RequirePackage{mathrsfs}

\newcommand{\cf}{cf.\@\xspace}
\newcommand{\resp}{resp.\@\xspace}

\newcommand{\al}{\alpha}
\newcommand{\bet}{\beta}
\newcommand{\ga}{\gamma}
\newcommand{\de}{\delta }

\newcommand{\f}{\varphi}
\newcommand{\h}{\eta}

\newcommand{\lam}{\lambda}

\newcommand{\om}{\omega}

\newcommand{\s}{\sigma}

\newcommand{\Clk}{\varGamma}
\newcommand{\D}{\varDelta}
\newcommand{\F}{\varPhi}
\newcommand{\Lam}{\varLambda}
\newcommand{\Om}{\varOmega}

\newcommand{\socc}{{\mc S_0}}

\newcommand{\msp[1]}[1]{\mspace{#1mu}}

\newcommand{\R}[1][n+1]{{\protect\mathbb R}^{#1}}

\newcommand{\Cc}{{\protect\mathbb C}}

\newcommand{\N}{{\protect\mathbb N}}

\newcommand{\eR}{\stackrel{\lower1ex \hbox{\rule{6.5pt}{0.5pt}}}{\msp[3]\R[]}}
\newcommand{\eN}{\stackrel{\lower1ex \hbox{\rule{6.5pt}{0.5pt}}}{\msp[1]\N}}
\newcommand{\eO}{\stackrel{\lower1ex \hbox{\rule{6pt}{0.5pt}}}{\msc O}}
\newcommand{\mf}[1]{\mathfrak {#1}}

\DeclareMathOperator{\id}{id}

\DeclareMathOperator{\Ad}{Ad}

\DeclareMathOperator{\tr}{tr}

\DeclareMathOperator{\diag}{diag}

\DeclareMathOperator{\card}{card}

\newcommand\im{\implies}
\newcommand\ra{\rightarrow}

\newcommand\hra{\hookrightarrow}

\newcommand\pa{\partial}
\newcommand\pde[2]{\frac {\partial#1}{\partial#2}}
 
\newcommand{\un}{\infty}
\newcommand{\A}{\forall}

\newcommand{\uu}{\cup}

\newcommand{\uuu}{\bigcup}

\newcommand{\uud}{ \stackrel{\lower 1ex \hbox {.}}{\uu}}
\newcommand{\uuud}[1]{ \stackrel{\lower 1ex \hbox {.}}{\uuu_{#1}}}
\newcommand\su{\subset}
\newcommand\Su{\Subset}

\newcommand{\sminus}[1][28]{\raise 0.#1ex\hbox{$\scriptstyle\setminus$}}

\newcommand{\ol}{\overline}

\newcommand{\wed}{\wedge}

\newcommand{\abs}[1]{\lvert#1\rvert}

\newcommand{\norm}[1]{\lVert#1\rVert}

\newcommand{\spd}[2]{\protect\langle #1,#2\protect\rangle}

\newcommand\cha[3]{{\bar\varGamma}_{#1#2}^#3}

\newcommand{\tit}{\textit}

\newcommand{\tup}{\textup}

\newcommand{\mc}{\protect\mathcal}
\newcommand{\msc}{\protect\mathscr}

\providecommand{\bysame}{\makebox[3em]{\hrulefill}\thinspace}

\newcommand{\bt}{\begin{thm}}
\newcommand{\bl}{\begin{lem}}
\newcommand{\bc}{\begin{cor}}
\newcommand{\bd}{\begin{definition}}
\newcommand{\bpp}{\begin{prop}}
\newcommand{\br}{\begin{rem}}
\newcommand{\bn}{\begin{note}}
\newcommand{\be}{\begin{ex}}
\newcommand{\bes}{\begin{exs}}
\newcommand{\bb}{\begin{example}}
\newcommand{\bbs}{\begin{examples}}
\newcommand{\ba}{\begin{axiom}}
\newcommand{\bas}{\begin{assumption}}

\newcommand{\et}{\end{thm}}
\newcommand{\el}{\end{lem}}
\newcommand{\ec}{\end{cor}}
\newcommand{\ed}{\end{definition}}
\newcommand{\epp}{\end{prop}}
\newcommand{\er}{\end{rem}}
\newcommand{\en}{\end{note}}
\newcommand{\ee}{\end{ex}}
\newcommand{\ees}{\end{exs}}
\newcommand{\eb}{\end{example}}
\newcommand{\ebs}{\end{examples}}
\newcommand{\ea}{\end{axiom}}
\newcommand{\eas}{\end{assumption}}

\newcommand{\bp}{\begin{proof}}
\newcommand{\ep}{\end{proof}}
\newcommand{\eps}{\renewcommand{\qed}{}\end{proof}}

\newcommand{\bal}{\begin{align}}

\newcommand{\bi}[1][1.]{\begin{enumerate}[\upshape #1]}
\newcommand{\bia}[1][(1)]{\begin{enumerate}[\upshape #1]}
\newcommand{\bin}[1][1]{\begin{enumerate}[\upshape\bfseries #1]}
\newcommand{\bir}[1][(i)]{\begin{enumerate}[\upshape #1]}
\newcommand{\bic}[1][(i)]{\begin{enumerate}[\upshape\hspace{2\cma}#1]}
\newcommand{\bis}[2][1.]{\begin{enumerate}[\upshape\hspace{#2\parindent}#1]}
\newcommand{\ei}{\end{enumerate}}

\newcommand\ndots{\raise 0.47ex \hbox {,}\hskip0.06em\cdots %
     \raise 0.47ex \hbox {,}\hskip0.06em} 

\newcommand{\q}{\quad}
\newcommand{\qq}{\qquad}

\newcommand{\hp}{\hphantom}

\newcommand\nd{\noindent}

\newskip\Csmallskipamount                                                
\Csmallskipamount=\smallskipamount
\newskip\Cmedskipamount
\Cmedskipamount=\medskipamount
\newskip\Cbigskipamount
\Cbigskipamount=\bigskipamount

\newcommand\cvs{\vspace\Csmallskipamount}   
\newcommand\cvm{\vspace\Cmedskipamount}

\newskip\csa
\csa=\smallskipamount

\newskip\cma
\cma=\medskipamount

\newskip\cba
\cba=\bigskipamount

\newdimen\spt
\newcommand\citem{\cvs\advance\itemno by
1{(\romannumeral\the\itemno})\hskip3pt}
\newcommand{\bitem}{\cvm\nd\advance\itemno by
1{\bf\the\itemno}\hspace{\cma}}

\newcount\itemno
\newcommand{\las}[1]{\label{S:#1}}

\newcommand{\lae}[1]{\label{E:#1}}
\newcommand{\lat}[1]{\label{T:#1}}
\newcommand{\lal}[1]{\label{L:#1}}
\newcommand{\lad}[1]{\label{D:#1}}

\newcommand{\lar}[1]{\label{R:#1}}

\newcommand{\rs}[1]{Section~\ref{S:#1}}

\newcommand{\rt}[1]{Theorem~\ref{T:#1}}
\newcommand{\rl}[1]{Lemma~\ref{L:#1}}

\newcommand{\re}[1]{\eqref{E:#1}}

\newcommand{\frl}[1]{Lemma~\ref{L:#1} on page~\tup{\pageref{L:#1}}}

\newcommand{\fre}[1]{\eqref{E:#1} on page~\tup{\pageref{E:#1}}}

\newskip\thmskip
\thmskip=\parindent

\newskip\hsk
\newenvironment{hinw}{\labelsep=0pt\begin{list}{}{\labelsep=0pt\itemindent=0pt\labelwidth=0pt\leftmargin=\parindent\rightmargin=0pt\partopsep=\cba}%
\item\it\nopagebreak\nopagebreak}%
{\end{list}}

\newcommand\bh{\begin{hinw}}
\newcommand{\eh}{\end{hinw}}

\newtheoremstyle{normal}
  {\cba}
  {\cba}
  {}
  {\thmskip}
  {\bfseries}
  {.}
  {\hsk}
  {}

\newtheoremstyle{abschnitt}
  {\cba}
  {\cba}
  {}
  {\thmskip}
  {\bfseries}
  {.}
  {\hsk}
  {}

\newtheoremstyle{italic}
  {\cba}
  {\cba}
  {\itshape}
  {\thmskip}
  {\bfseries}
  {.}
  {\hsk}
  {}

\newtheoremstyle{aufgaben}
  {\cba}
  {\cba}
  {}
  {}
  {\normalsize\bfseries}
  {.}
  {\hsk}
  {}

\newtheoremstyle{break}
  {\cba}
  {\cba}
  {\itshape}
  {}
  {\bfseries}
  {.}
  {\newline}
  {}

\theoremstyle{italic}
\newtheorem{thm}[subsection]{Theorem}
\newtheorem{lem}[subsection]{Lemma}
\newtheorem{prop}[subsection]{Proposition}
\newtheorem{cor}[subsection]{Corollary}

\theoremstyle{normal}
\newtheorem{rem}[subsection]{Remark}
\newtheorem{definition}[subsection]{Definition}
\newtheorem{example}[subsection]{Example}
\newtheorem{examples}[subsection]{Examples}
\newtheorem{ex}[subsection]{Exercise}
\newtheorem{note}[subsection]{}
\newtheorem{axiom}[subsection]{Axiom}
\newtheorem{assumption}[subsection]{Assumption}

\theoremstyle{aufgaben}
\newtheorem{exs}[subsection]{Exercises}

\numberwithin{equation}{section}
\numberwithin{figure}{section}

\newenvironment{textequation}[1][0.8]
{\begin{equation}
\begin{aligned}
\begin{minipage}{#1\linewidth}}
{\end{minipage}
\end{aligned}
\end{equation}
\ignorespacesafterend}

\newcommand{\btext}{\begin{textequation}}
\newcommand{\etext}{\end{textequation}}

\def\hinweis{\@startsection{subsection}{2}%
 \z@{0.7\linespacing\@plus 0.5\linespacing}{0.7\linespacing}%
{\normalfont\itshape\indent}}

\newcounter{hours}\newcounter{minutes}
\newcommand{\printtime}{%
\setcounter{hours}{\time/60}%
\setcounter{minutes}{\time-\value{hours}*60}%
\ifthenelse{\value{minutes}<10}{\thehours :0\theminutes}{\thehours:\theminutes}}

\usepackage[german,english]{babel}
\usepackage{graphicx}
\RequirePackage{amsmath}
\RequirePackage{bm}
\RequirePackage{amssymb}
\RequirePackage{upref}
\RequirePackage{amsthm}
\RequirePackage{enumerate}
\RequirePackage{pb-diagram}
\RequirePackage{amsfonts}
\RequirePackage[mathscr]{eucal}
\RequirePackage{verbatim}
\RequirePackage{xr}
\RequirePackage{graphicx}
\usepackage{calc}
\usepackage{xspace}
\usepackage{dsfont}

\makeatletter
\RequirePackage{color}
\newcommand{\ann}[1]{\renewcommand{\@makefnmark}{\mbox{$^{\color{red}{\@thefnmark}}$}}%
\footnote {#1}}
\makeatother

\newcommand{\rmann}{\renewcommand{\ann}[1]{}}

\RequirePackage{upref}
\RequirePackage{amsthm}
\RequirePackage{enumerate}
\usepackage[mathscr]{eucal}

\usepackage{xr-hyper}

\usepackage{calc}

\newlength{\oddsidemarginlength}
\newlength{\topmarginlength}

\newcounter{numberoflines}
\newcounter{tempcc}
\oddsidemargin=\oddsidemarginlength
\evensidemargin=\oddsidemargin
\topmargin=\topmarginlength
\usepackage[colorlinks=true,linkcolor=blue,citecolor=blue,urlcolor=blue]{hyperref}  

\begin{document}

\flushbottom


\title[A unified quantum theory]{ A unified quantization of gravity and other fundamental forces of nature}

\author{Claus Gerhardt}
\address{Ruprecht-Karls-Universit\"at, Institut f\"ur Angewandte Mathematik,
Im Neuenheimer Feld 295, 69120 Heidelberg, Germany}
\email{\href{mailto:gerhardt@math.uni-heidelberg.de}{gerhardt@math.uni-heidelberg.de}}
\urladdr{\href{http://www.math.uni-heidelberg.de/studinfo/gerhardt/}{http://www.math.uni-heidelberg.de/studinfo/gerhardt/}}

%

\date{\today}
%

\dedicatory{Dedicated to Robert Finn on the occasion of his 100th birthday}

\rmann
\begin{abstract} 
We quantize the interaction of gravity with Yang-Mills and spinor fields, hence offering a quantum theory incorporating all four fundamental forces of nature. Let us\ann{as} abbreviate the spatial Hamilton functions of the Standard Model by $H_{SM}$ and the Hamilton function of gravity by $H_G$. Working in a fiber bundle $E$ with base space $\socc=\R[n]$, where the fiber elements are Riemannian metrics, we can express the Hamilton functions in the form $H_G+H_{SM}=H_G+t^{-\frac23}\tilde H_{SM}$ if $n=3$, where $\tilde H_{SM}$ depends on metrics $\s_{ij}$ satisfying $\det{\s_{ij}}=1$. In the quantization process, we quantize $H_G$ for general $\s_{ij}$ but $\tilde H_{SM}$ only for $\s_{ij}=\de_{ij}$ by the usual methods of QFT. Let $v$ \resp $\psi$ be the spatial eigendistributions of the respective Hamilton operators, then, the solutions $u$ of the Wheeler-DeWitt equation are given by $u=wv\psi$, where $w$ satisfies an ODE and $u$ is evaluated at $(t,\de_{ij})$ in the fibers.
\end{abstract}

\maketitle

\tableofcontents

\setcounter{section}{0}
\section{Introduction}
General relativity is a Lagrangian theory, i.e., the Einstein equations are derived as the Euler-Lagrange equation of the Einstein-Hilbert functional 
\begin{equation}
\int_N(\bar R-2\Lam),
\end{equation}
where $N=N^{n+1}$, $n\ge 3$, is a globally hyperbolic Lorentzian manifold, $\bar R$ the scalar curvature and $\Lam$ a cosmological constant. We also omitted the integration density in the integral. In order to apply a Hamiltonian description of general relativity, one usually defines a time function $x^0$ and considers the foliation of $N$ given by the slices
\begin{equation}
M(t)=\{x^0=t\}.
\end{equation}
We may, without loss of generality, assume that the spacetime metric splits
\begin{equation}\lae{1.3}
d\bar s^2=-w^2(dx^0)^2+g_{ij}(x^0,x)dx^idx^j,
\end{equation}
\cf \cite[Theorem 3.2]{cg:qgravity}. Then, the Einstein equations also split into a tangential part
\begin{equation}
G_{ij}+\Lam g_{ij}=0
\end{equation}
and a normal part
\begin{equation}
G_{\al\bet}\nu^\al\nu^\bet-\Lam=0,
\end{equation}
where the naming refers to the given foliation. For the tangential Einstein equations one can define equivalent Hamilton equations due to the groundbreaking paper by Arnowitt, Deser and Misner \cite{adm:old}. The normal Einstein equations can be expressed by the so-called Hamilton condition
\begin{equation}\lae{1.6}
\mc H=0,
\end{equation}
where $\mc H$ is the Hamiltonian used in defining the Hamilton equations. In the canonical quantization of gravity the Hamiltonian is transformed  to a partial differential operator of hyperbolic type $\hat{\mc H}$ and the possible quantum solutions of gravity are supposed to satisfy the so-called Wheeler-DeWitt equation
\begin{equation}\lae{1.7}
\hat{\mc H}u=0
\end{equation}
in an appropriate setting, i.e., only the Hamilton condition \re{1.6} has been quantized, or equivalently, the normal Einstein equation, while the tangential Einstein equations have been ignored.

In \cite{cg:qgravity} we solved the equation \re{1.7} in a fiber bundle $E$ with base space $\socc$,
\begin{equation}
\socc=\{x^0=0\}\equiv M(0),
\end{equation}
and fibers $F(x)$, $x\in\socc$,
\begin{equation}
F(x)\su T^{0,2}_x(\socc),
\end{equation}
the elements of which are the positive definite symmetric tensors of order two, the Riemannian metrics in $\socc$. The hyperbolic operator $\hat{\mc H}$ is then expressed in the form 
\begin{equation}\lae{1.10}
\hat{\mc H}=-\D-(R-2\Lam)\f,
\end{equation}
where $\D$ is the Laplacian of the DeWitt metric given in the fibers, $R$ the scalar curvature of the metrics $g_{ij}(x)\in F(x)$, and $\f$ is defined by
\begin{equation}\lae{1.11}
\f^2=\frac{\det g_{ij}}{\det\rho_{ij}},
\end{equation}
where $\rho_{ij}$ is a fixed metric in $\socc$ such that instead of densities we are considering functions. The Wheeler-DeWitt equation could be solved in $E$ but only as an abstract hyperbolic equation. The solutions could not be split in corresponding spatial and temporal eigenfunctions.

The underlying mathematical reason for the difficulty was the presence of the term $R$ in the quantized equation, which prevents the application of separation of variables, since the metrics $g_{ij}$ are the spatial variables. In a recent paper \cite{cg:qgravity3} we overcame this difficulty by quantizing the Hamilton equations instead of the Hamilton condition. 
 
 As a result we obtained the equation
 \begin{equation}\lae{1.12}
-\D u=0
\end{equation}
in $E$, where the Laplacian is the Laplacian in \re{1.10}. The lower order terms of $\hat{\mc H}$ 
\begin{equation}
(R-2\Lam)\f
\end{equation}
were eliminated during the quantization process. However, the equation \re{1.12} is only valid provided $n\not=4$, since the resulting equation actually looks like
\begin{equation}
-(\frac n2-2)\D u=0.
\end{equation}
This restriction seems to be acceptable, since $n$ is the dimension of the base space $\socc$ which, by general consent, is assumed to be $n=3$. The fibers add additional dimensions to the quantized problem, namely,
\begin{equation}
\dim F=\frac {n(n+1)}2\equiv m+1.
\end{equation}
The fiber metric, the DeWitt metric, which is responsible for the Laplacian in \re{1.12} can be expressed in the form
\begin{equation}\lae{1.16.1}
ds^2=-\frac{16(n-1)}n dt^2+\f G_{AB}d\xi^Ad\xi^B,
\end{equation}
where the coordinate system is
\begin{equation}\lae{1.17}
(\xi^a)= (\xi^0,\xi^A)\equiv (t,\xi^A).
\end{equation}
The $(\xi^A)$, $1\le A\le m$, are coordinates for the hypersurface
\begin{equation}\lae{1.18}
M\equiv M(x)=\{(g_{ij}):t^4=\det g_{ij}(x)=1,\A\, x\in\socc\}.
\end{equation}
We also assumed that $\socc=\R[n]$ and that  the metric $\rho_{ij}$ in \re{1.11} is the Euclidean metric $\de_{ij}$. It is well-known that $M$ is a symmetric space
\begin{equation}\lae{1.19.1}
M=SL(n,\R[])/SO(n)\equiv G/K.
\end{equation}
It is also easily verified that the induced metric of $M$ in $E$ is isometric to the Riemannian metric of the coset space $G/K$.

Now, we were in a position to use separation of variables, namely, we wrote a solution of \re{1.12} in the form 
\begin{equation}
u=w(t)v(\xi^A),
\end{equation}
where $v$ is a spatial eigenfunction of the induced Laplacian of $M$ 
\begin{equation}\lae{1.21}
-\D_Mv\equiv -\D v=(\abs\lam^2+\abs\rho^2)v
\end{equation}
and $w$ is a temporal eigenfunction satisfying the ODE
\begin{equation}\lae{1.22}
\Ddot w+m t^{-1}\dot w+\mu_0 t^{-2}w=0
\end{equation}
with
\begin{equation}
\mu_0=\frac{16(n-1)}n(\abs \lam^2+\abs\rho^2).
\end{equation}

The eigenfunctions of the Laplacian in $G/K$ are well-known and we chose the kernel of the Fourier transform in $G/K$ in order to define the eigenfunctions. This choice also allowed us to use Fourier quantization similar to the Euclidean case such that the eigenfunctions are transformed to Dirac measures and the Laplacian to a multiplication operator in Fourier space.

In the present paper we like to quantize the Einstein-Hilbert functional combined with the functionals of the other fundamental forces of nature, i.e., we look at  the Lagrangian functional
\begin{equation}\lae{1.1}
\begin{aligned}
J&=\al_N^{-1}\int_{\tilde\Om}(\bar R-2\Lam)-\int_{\tilde\Om}\tfrac14 \ga_{\bar a\bar b}\bar g^{\mu\rho_2}\bar g^{\lam\rho_1}F^{\bar a}_{\mu\rho_1}F^{\bar b}_{\rho_2\lam}\\
&\hp =\;\;\,-\int_{\tilde\Om}\{\tfrac12 \bar g^{\mu\lam}\ga_{\bar a\bar b}\F^{\bar a}_\mu\bar\F^{\bar b}_\lam+V(\F)\}\\
&\hp =\;\;+\int_{\tilde\Om}\{\tfrac12[\tilde\psi_{I}E^\mu_a\ga^a(D_\mu \psi)^{I}+\overline{\tilde\psi_{I}E^\mu_a\ga^a(D_\mu \psi)^{I}}]+m\tilde\psi_{I}\psi^{I}\},
\end{aligned}
\end{equation}
where $\al_N$ is a positive coupling constant,  $\tilde\Om\Su N=N^{n+1}$ and $N$ a globally hyperbolic spacetime with metric $\bar g_{\al\bet}$, $0\le \al,\bet\le  n$, where the metric splits as in \re{1.3}.

The functional $J$ consists of the Einstein-Hilbert functional, the Yang-Mills and Higgs functional and a massive Dirac term. 

The Yang-Mills field $(A_\mu)$
\begin{equation}
A_\mu=f_{\bar c}A^{\bar c}_\mu
\end{equation}
corresponds to the adjoint representation of a compact, semi-simple Lie group $\mc G$ with Lie algebra $\mf g$. The $f_{\bar c}$,
\begin{equation}
f_{\bar c}=(f^{\bar a}_{\bar c\bar b})
\end{equation}
are the structural constants of $\mf g$.

We assume the Higgs field $\F=(\F^{\bar a})$ to have complex valued components.

The spinor field $\psi=(\psi^I_A)$ has  a spinor index $A$, $1\le A\le n_1$, and a colour index $I$, $1\le I\le n_2$. Here, we suppose that the Lie group has a unitary representation $R$ such that
\begin{equation}
t_{\bar c}=R(f_{\bar c})
\end{equation}
are antihermitian matrices acting on $\Cc^{n_2}$. The symbol $A_\mu\psi$ is now defined by
\begin{equation}
A_\mu\psi=t_{\bar c}\psi A^{\bar c}_\mu.
\end{equation}

There are some major difficulties in achieving a quantization of the functional in \re{1.1}. Quantizing the Hamilton equations, to avoid the problem with the scalar curvature term, runs into technical difficulties, even if the required quantization  of the matter fields in  curved spacetimes could be achieved, since the resulting operator would no longer be hyperbolic because the  elliptic parts of the gravitational \resp matter Hamiltonians would have different signs in case $n=3$. This particular problem would not occur when the Hamilton condition would be quantized. The Hamilton condition  has the form
\begin{equation}\lae{1.29.1}
H_G+H_{YM}+H_D+H_H=0,
\end{equation}
where the subscripts refer to gravity, Yang-Mills, Dirac and Higgs. On the left-hand side are the Hamilton functions of the respective fields. They depend on the  Riemannian metrics $g_{ij}$, the Yang-Mills connections and the spinor and Higgs fields. The main part of the quantized gravitational Hamiltonian is a second order hyperbolic differential operator with respect to the variables $g_{ij}$ while the scalar curvature term $R$ is of zero order. Having this in mind we also apply these categories to the gravitational Hamilton function where the main part, quadratic in the conjugate momenta, is said to be of second order and the zero order terms consist of the scalar curvature and the cosmological constant $\Lam$. Similarly we consider the matter Hamilton functions to be zero order terms with respect to the metric $g_{ij}$, i.e., there is no qualitative difference by assuming $g_{ij}$ to be flat or non-flat, or more precisely, quantizing a matter Hamiltonian in a curved spacetime when $g_{ij}$ is a given, fixed metric and not a variable is qualitatively the same as quantizing it for  the Euclidean metric, though the task is certainly more difficult.

Thus, the difficulties arising by quantizing the Hamilton condition can best be explained by considering the Wheeler-DeWitt equation
\begin{equation}
\hat H_Gu=0\qq \text{in } E,
\end{equation}
\cf  \re{1.6}, where we wrote $\hat{\mc H}$ instead of $\hat H_G$. This is a hyperbolic differential equation which can be expressed by
\begin{equation}
\hat H_Gu= -\D u+\f (R-2\Lam)u=0,
\end{equation}
where the Laplacian is the Laplacian of the fiber metric \re{1.16.1}. In the coordinate system \re{1.17} we get
\begin{equation}\lae{1.32.1}
\hat H_Gu=t^{-m}\frac{\pa}{\pa t}(t^m\frac{\pa u}{\pa t})-t^{-2}\D_Mu+t^2(R-2\Lam)u,
\end{equation}
where $M$ is the hypersurface \re{1.18}. Since $M$ is isometric to the symmetric space \re{1.19.1} it is mathematically irresistible to solve \re{1.32.1} by applying separation of variables and using 
the functions of the Fourier kernel of $M$ as spatial eigenfunctions $v$, where $v=v(\s_{ij})$, $\s_{ij}$ are the elements of $M$. Since
\begin{equation}\lae{1.33.1}
g_{ij}(x)=t^{\frac4n}\s_{ij}(x)
\end{equation}
the critical term $R$ can be expressed as
\begin{equation}
R(g_{ij})=t^{-\frac4n}R(\s_{ij})
\end{equation}
due to the relation between the scalar curvatures of conformal metrics. 

Thus, it is obvious that the ansatz
\begin{equation}
u=wv,
\end{equation}
where $w=w(t)$ solves an ODE is only possible if $R(\s_{ij})$ is constant
\begin{equation}\lae{1.36.1}
R(\s_{ij})=\lam_0.
\end{equation}
The constant is arbitrary but of course determined by the metrics we are considering to be important, e.g., in case of a black hole we would choose $\s_{ij}$  to be the limit metric of a converging sequence of Cauchy hypersurfaces of the interior region of the black hole which converge to the event horizon topologically but the induced metrics of which converge to a Riemannian metric, \cf  \cite{cg:qbh,cg:qbh2} or \cite[Chapters 4 \& 5]{cg:qgravity-book}. In the present case, where we want to include the matter fields of the Standard Model we could choose $\s_{ij}=\de_{ij}$.

However, this ansatz implies that the Wheeler-DeWitt equation is not solved for all $(t,\s_{ij})$ but only for the $\s_{ij}$ satisfying \re{1.36.1}. Given the simplicity and mathematical beauty of the solution, we are inclined to accept this restriction.

Let us now consider the quantization of the Hamilton condition \re{1.29.1} taking all Hamilton functions into account. In view of the relation \re{1.33.1} let me propose the following model:
If we were able to express the non-gravitational Hamiltonians as
\begin{equation}\lae{1.32}
H_{YM}=t^{ p} \tilde H_{YM}, \q H_D=t^p\tilde H_D,\q H_H=t^p\tilde H_H,
\end{equation}
where the embellished Hamiltonians depend on $\s_{ij}$, then, by choosing in addition $n=3$ and $\s_{ij}=\de_{ij}$, these Hamiltonians could be quantized by the known methods of QFT, if the Lie groups would be chosen appropriately. The Wheeler-DeWitt equation would then not be solved for all $(t,\s_{ij})$ but only for $(t,\de_{ij})$. However, the spatial eigendistributions of the Hamilton operator $\hat H_G$, i.e., the  eigendistributions of the Laplacian of $M$, \cf \re{1.21}, would still be used but they would be evaluated at $\s_{ij}=\de_{ij}$.

In \rs{4} we shall prove that the expressions in \re{1.32} are indeed valid with $p=-\frac23$ provided $n=3$ and provided that the mass term\ann{the} in the Dirac Lagrangian and the Higgs Lagrangian are slightly modified. The embellished Hamiltonians are then standard Hamiltonians without any modifications, for details we refer to \rs{4}. The Hamilton constraint then has the form
\begin{equation}
\begin{aligned}
H&=H_G+H_{YM}+H_H+H_D\\
&=H_G+t^{-\frac23}(\tilde H_{YM}+\tilde H_H+\tilde H_D)\\
&\equiv H_G+t^{-\frac23} \tilde H_{SM}=0,
\end{aligned}
\end{equation}
where the subscript $SM$ refers to the fields  of the Standard Model or to a  corresponding subset of fields. The solutions of the Wheeler-DeWitt equation
\begin{equation}
\hat Hu=0
\end{equation}
can then be achieved by using separation of variables. We proved:
\bt\lat{1.1}
Let $n=3$, $v=e_{\lam,b_0}$ and let $\psi$ be an  eigendistribution of $\tilde H_{SM}$ when $\s_{ij}=\de_{ij}$ such that
\begin{equation}
- \D_M e_{\lam,b_0}=(\abs\lam^2+1)e_{\lam,b_0},
\end{equation}
\begin{equation}
\tilde H_{SM}\psi=\lam_1\psi,\qq\lam_1\ge 0,
\end{equation}
and let $w$ be a solution of the  ODE 
\begin{equation}\lae{4.147.1}
\begin{aligned}
t^{-m}\pde{}{t}(t^m \pde wt)&+\frac{32}3 (\abs\lam^2+1)t^{-2}w+\frac{32}3\al_N^{-1}\lam_1 t^{-\frac23}w\\
& \q+\frac{64}3\al_N^{-2}\Lam t^2w=0
\end{aligned}
\end{equation}
then
\begin{equation}
u=w e_{\lam,b_0}\psi
\end{equation}
is a solution of the Wheeler-DeWitt equation
\begin{equation}
\hat Hu=0,
\end{equation}
where $e_{\lam,b_0}$ is evaluated at $\s_{ij}=\de_{ij}$ and where we note that $m=5$.
\et
We shall refer to $e_{\lam,b_0}$ and $\psi$ as the spatial eigenfunctions and to $w$ as the temporal eigenfunction.

\br
We could also apply the respective Fourier transforms to $-\tilde \D e_{\lam,b_0}$ \resp $\tilde H_{SM}\psi$ and consider
\begin{equation}
w\hat e_{\lam,b_0}\hat\psi
\end{equation}
as the solution in Fourier space, where $\hat\psi$ would be expressed with the help of the ladder operators.
\er
The temporal eigenfunctions are analyzed in \rs{5}. They must satisfy an ODE of the form
\begin{equation}\lae{5.2.1}
\Ddot w+5 t^{-1} \dot w+m_1 t^{-2}w+m_2^2 t^{-\frac23}w+m_3 t^2w=0,
\end{equation}
where
\begin{equation}\lae{5.3.1}
m_1\ge\frac{32}3,\q m_2\ge 0,\q m_3\in\R[].
\end{equation}
For simplicity we shall only state the result when $m_3=0$ which is tantamount to setting $\Lam=0$. 
\bt 
Assume $m_3=0$ and $m_2>0$, then the solutions of the ODE \re{5.2.1} are generated by
\begin{equation}
J(\tfrac32\sqrt{m_1-4}\,i,\tfrac32 m_2 t^\frac23) t^{-2}
\end{equation}
and
\begin{equation}
J(-\tfrac32\sqrt{m_1-4}\,i,\tfrac32 m_2 t^\frac23) t^{-2},
\end{equation}
where $J(\lam,t)$ is the Bessel function of the first kind.
\et
\bl\lal{5.2.1}
The solutions in the theorem above diverge to complex infinity if $t$ tends to zero and they converge to zero if $t$ tends to infinity.
\el

\section{Definitions and notations} 
Greek indices $\al$, $\bet$ range from $0$ to $n$, Latin $i,j,k$ from $1$ to $n$ and we stipulate $0\le a, b\le n$ but $1\le a', b'\le n$. Barred indices $\bar a$ refer to the Lie algebra $\mf g$, $1\le \bar a\le n_0=\dim\mf g$.

$\ga_{\bar a\bar b}$ is the Cartan-Killing metric.

The Dirac matrices are denoted by $\ga^a$ and they satisfy
\begin{equation}
\ga^a\ga^b+\ga^b\ga^a=2\h^{ab}I,
\end{equation}
where $\h_{ab}$ is the Minkowski metric with signature $(-,+,\ldots,+)$. $\ga^0$ is antihermitian and $\ga^{a'}$ Hermitian.

The indices $a,b$ are always raised or lowered with the help of the Minkowski metric, Greek indices with the help of the spacetime metric $\bar g_{\al\bet}$.

The $\ga^a$ act in
\begin{equation}
\Cc^{2^{\frac{n+1}2}},
\end{equation}
if $n$ is odd and in
\begin{equation}
\Cc^{2^\frac n2}\oplus \;\Cc^{2^\frac n2},
\end{equation}
if $n$ is even. In both cases we simply refer to these spaces as
\begin{equation}
\Cc^{n_1},
\end{equation}
i.e., the spinor index $A$ has range $1\le A\le n_1$.

The colour index $I$ has range $1\le I\le n_2$ and hence a spinor field $\psi^I_A$ has values in
\begin{equation}
\Cc^{n_1}\otimes \Cc^{n_2}.
\end{equation}

Finally, a Hermitian form $\spd\cdot\cdot$ is antihermitian in the first argument.

\section{Spinor fields}\las{3}
The Lagrangian of the spinor field is stated in \re{1.1}. Here, $\psi=(\psi^I_A)$ is a multiplet of spinors with spin $\tfrac12$; $A$ is the spinor index, $1\le A\le n_1$, and $I$, $1\le I\le n_2$, the \tit{colour} index. 
We shall also lower or raise the index $I$ with the help of the Euclidean metric $(\de_{IJ})$.

Let $\Clk_\mu$ be the spinor connection
\begin{equation}
\Clk_\mu=\tfrac14 \om_{\mu\hp ba}^{\hp\mu b}\ga_b\ga^a,
\end{equation}
then the covariant derivative $D_\mu\psi$ is defined by 
\begin{equation}
D_\mu\psi=\psi_{,\mu}+\Clk_\mu\psi+A_\mu\psi.
\end{equation}

Let $(e^b_\lam)$ be a $n$-bein such that
\begin{equation}
\bar g_{\mu\lam}=\h_{ab}e^a_\mu e^b_\lam,
\end{equation}
where $(\h_{ab})$ is the Minkowski metric, and let $(E^\mu_a)$ be its inverse 
\begin{equation}
E^\mu_a=\h_{ab}\bar g^{\mu\lam}e^b_\lam,
\end{equation}
\cf \cite[p. 246]{eguchi:book}. 

The covariant derivative of $E^\al_a$ with respect to $(\bar g_{\al\bet})$ is then given by
\begin{equation}
E^\al_{a;\mu}=E^\al_{a,\mu}+\cha \mu\bet\al E^\bet_a
\end{equation}
and
\begin{equation}
\om_{\mu\hp ba}^{\hp\mu b}=E^\lam_{a;\mu}e^b_\lam=-E^\lam_ae^b_{\lam;\mu},
\end{equation}
hence the spin connection $\Clk_\mu$ can be expressed as
\begin{equation}
\Clk_\mu=\tfrac14 \om_{\mu\hp ba}^{\hp\mu b}\ga_b\ga^a=\tfrac14 E^\lam_{a;\mu}e^b_\lam \ga_b\ga^a=-\tfrac14 E^\lam_ae^b_{\lam;\mu}\ga_b\ga^a. 
\end{equation}

We shall first show:
\bl\lal{3.1}
Let $\bar g_{\al\bet}$ be a fixed spacetime metric that is split by the time function $x^0$, then there exists an orthonormal frame $(e^a_\lam)$ such that
\begin{equation}
e^0_k=0,\qq 1\le k\le n,
\end{equation}
and
\begin{equation}
e^{a'}_{k;0}=e^{a'}_{,0}-\cha k0\lam e^{a'}_\lam=0
\end{equation}
for all $1\le a'\le n$ and $1\le k\le n$.
\el
\bp
Assume that
\begin{equation}
\bar g_{00}=-w^2,
\end{equation}
then define the conformal metric
\begin{equation}
\tilde g_{\al\bet}=w^{-2}\bar g_{\al\bet}.
\end{equation}
The curves
\begin{equation}
(\ga^\al(t,x))=(t,x^i),\qq x\in\socc,
\end{equation}
are then geodesics with respect to $\tilde g_{\al\bet}$. Let $(\hat e^{a'}_\lam)$, $1\le a'\le n$, be an orthonormal frame in $T^{0,1}(\socc)\hra T^{0,1}(N)$ such that
\begin{equation}
\hat e^{a'}_0=0\qq\A \, 1\le a'\le n.
\end{equation}
The $\hat e^{a'}$ depend on $x=(x^i)\in \socc$. Let $(\tilde e^{a'}_\lam)(t,x)$ be the solutions of the flow equations
\begin{equation}
\begin{aligned}
\frac D{dt}\tilde e^{a'}_\lam&=0,\\
\tilde e^{a'}_\lam(0,x)&=\hat e^{a'}_\lam(x),
\end{aligned}
\end{equation}
i.e., we parallel transport $\hat e^{a'}$ along the geodesics. Setting
\begin{equation}
(\tilde e^0_\lam)=(1,0,\ldots,0)
\end{equation}
the $(\tilde e^a_\lam)$ are then an orthonormal frame of $1$-forms in $(N,\tilde g_{\al\bet})$ such that the $\tilde e^a$ satisfy
\begin{equation}\lae{3.16}
\tilde e^a_{\lam:0}=0\qq\A\, 0\le a\le n,
\end{equation}
where we indicate covariant differentiation with respect to $\tilde g_{\al\bet}$ by a colon.

Define $e^a_\lam$ by
\begin{equation}
e^a_\lam=w \tilde e^a_\lam,
\end{equation}
then the $e^a_\lam$ are orthonormal frames in $(N,\bar g_{\al\bet})$. The Christoffel symbols $\cha \al\bet\ga$ \resp $\tilde \Clk^\ga_{\al\bet}$ are related by the formula
\begin{equation}\lae{3.18}
\begin{aligned}
\cha \al\bet\ga= \tilde \Clk^\ga_{\al\bet} - w^{-1} w_\al \de^\ga_\bet +w^{-1}w_\bet \de^\ga_\al -w^{-1} \check w^\ga \tilde g_{\al\bet},
\end{aligned}
\end{equation}
where 
\begin{equation}
\check w^\ga=\tilde g^{\ga\lam}w_\lam.
\end{equation}

In view of \re{3.16} we then infer
\begin{equation}
0=\tilde e^{a'}_{j:0}=\dot{\tilde e}^{a'}_j-\tilde \Clk^k_{0j}\tilde e^{a'}_k
\end{equation}
and we deduce further
\begin{equation}
\begin{aligned}
e^{a'}_{j;0}&=\dot w\tilde e^{a'}_j+w\dot{\tilde e}^{a'}_j-\cha 0jk w \tilde e^{a'}_k\\
&=\dot w \tilde e^{a'}_j+\tilde\Clk ^k_{0j}w\tilde e^{a'}_k -\cha 0jk w\tilde e^{a'}_k\\
&=0
\end{aligned}
\end{equation}
because of \re{3.18}.
\ep
Subsequently we shall always use these particular orthonormal frames.

We are now able to simplify the expressions for the spin connections
\begin{equation}
\Clk_\mu= -\tfrac14 E^\lam_a e^b_{\lam;\mu}\ga_a\ga^b.
\end{equation}
We have
\begin{equation}
\begin{aligned}
4\Clk_0&=-E^\lam_a e^b_{\lam;0}\ga_b\ga^a\\
&=-E^\lam_ae^0_{\lam;0}\ga_0\ga^a-E^\lam_ae^{b'}_{\lam;0}\ga_{b'}\ga^a\\
&=-E^0_0 e^0_{0;0}\ga_0\ga^0-E^i_{a'}e^0_{i;0}\ga_0\ga^{a'}-E^0_0e^{b'}_{0;0}\ga_{b'}\ga^0-E^i_{a'}e^{b'}_{i;0}\ga_{b'}\ga^{a'}\\
&=-E^i_{a'}e^0_{i;0}\ga_0\ga^{a'}-E^0_0e^{b'}_{0;0}\ga_{b'}\ga^0
\end{aligned}
\end{equation}
in view of \rl{3.1} and the fact that
\begin{equation}
e^0_{0;0}=0.
\end{equation}

The matrices $\ga_0\ga^{a'}$ and $\ga_{b'}\ga^0$ are hermitian, since $\ga^0$ is antihermitean, $\ga^{a'}$ hermitean and there holds
\begin{equation}
\ga_0\ga^{a'}=-\ga^{a'}\ga_0.
\end{equation}

Hence, the quadratic form
\begin{equation}
\tilde\psi E^0_a \ga^a\Clk_0\psi=-iE^0_0\bar\psi\Clk_0\psi
\end{equation}
is imaginary and will be eliminated by adding its complex conjugate. $\Clk_0$ can therefore be ignored which we shall indicate by writing
\begin{equation}
\Clk_0\simeq 0.
\end{equation}
A similar notation should apply to other terms that will be cancelled when adding the complex conjugates.

Let us consider $\Clk_k$:
\begin{equation}
\begin{aligned}
4\Clk_k&=-E^\lam_a e^b_{\lam;k}\ga_b\ga^a\\
&=-E^\lam_ae^0_{\lam;k}\ga_0\ga^a-E^\lam_ae^{b'}_{\lam;k}\ga_{b'}\ga^a\\
&=-E^0_0 e^0_{0;k}\ga_0\ga^0-E^i_{a'}e^0_{i;k}\ga_0\ga^{a'}-E^0_0e^{b'}_{0;k}\ga_{b'}\ga^0-E^i_{a'}e^{b'}_{i;k}\ga_{b'}\ga^{a'}.
\end{aligned}
\end{equation}
The first term on the right-hand side vanishes, since
\begin{equation}
e^0_{0;k}=w_k-\cha0k0w=0.
\end{equation}
Furthermore, there holds
\begin{equation}
e^0_{i;k}=-\cha ik0w=-\tfrac12\dot g_{ik}w^{-1}
\end{equation}
and
\begin{equation}
e^{b'}_{0;k}=-\cha 0kje^{b'}_j=-\tfrac12 g^{lj}\dot g_{kl}e^{b'}_j,
\end{equation}
yielding
\begin{equation}\lae{3.32}
\begin{aligned}
4\Clk_k&=\tfrac12 \dot g_{ik}w^{-1}E^i_{a'}\ga_0\ga^{a'}+\tfrac12 w^{-1}g^{lj}\dot g_{kl} e^{b'}_i\ga_{b'}\ga^0-E^i_{a'}e^{b'}_{i;k}\ga_{b'}\ga^{a'}\\
&=w^{-1}\dot g_{ik}E^i_{a'}\ga_0\ga^{a'}-E^i_{a'}e^{b'}_{i;k}\ga_{b'}\ga^{a'},
\end{aligned}
\end{equation}
since
\begin{equation}
\ga_0\ga^{a'}=-\ga^{a'}\ga_0.
\end{equation}
The first term on the right-hand side of \re{3.32} has to be eliminated because of the presence of $\dot g_{ik}$. To achieve this fix a Riemannian metric $\rho_{ij}=\rho_{ij}(x)\in T^{0,2}(\socc)$ and define the function $\f$ by
\begin{equation}\lae{3.34.1} 
\f=\sqrt{\frac{\det g_{ij}}{\det \rho_{ij}}}
\end{equation}
and the spinors $\chi=(\chi^i_A)$ by
\begin{equation}\lae{3.35}
\chi=\sqrt\f\psi,
\end{equation}
then
\begin{equation}
\dot \chi=\sqrt\f\dot\psi+\tfrac14 g^{ij}\dot g_{ij}\chi
\end{equation}
and
\begin{equation}
\chi_{,k}=\tfrac12 \f_k\f^{-1/2}\chi +\sqrt\f \psi_{,k}.
\end{equation}
Looking at the real part of the quadratic form
\begin{equation}
i\tilde \chi E^k_{a'}\ga^{a'}\chi_{,k}
\end{equation}
we deduce that
\begin{equation}
\chi_{,k}\simeq \sqrt\f \psi_{,k}.
\end{equation}

Moreover, we infer
\begin{equation}
\begin{aligned}
i\tilde\psi E^k_{c'}\ga^{c'}\Clk_k\psi&= i\bar \psi E^k_{c'}\ga^0\ga^{c'}\Clk_k\psi\\
&=\tfrac14 i\bar\psi E^k_{c'}E^j_{a'}w^{-1}\dot g_{jk}\ga^0\ga^{c'}\ga_0\ga^{a'}\psi\\
&\q -\tfrac14 i \bar\psi E^k_{c'}E^j_{a'}e^{b'}_{j;k}\ga^0\ga^{c'}\ga_{b'}\ga^{a'}\psi.
\end{aligned}
\end{equation}
We now observe that
\begin{equation}
\ga^0\ga^{c'}\ga_0\ga^{a'}=-\ga^0\ga_0\ga^{c'}\ga^{a'}=-\ga^{c'}\ga^{a'},
\end{equation}
hence
\begin{equation}
E^k_{c'}E^j_{a'}\ga^0\ga^{c'}\ga_0\ga^{a'}=-E^k_{c'}E^j_{a'}\ga^{c'}\ga^{a'}=-g^{jk}
\end{equation}
and we conclude
\begin{equation}
\begin{aligned}
i\tilde\psi E^\mu_c\ga^cD_\mu\psi\f&\simeq-i\bar\chi \dot\chi w^{-1}\\
&\q +i\bar\chi E^k_{c'}\ga^0\ga^{c'}\{\chi_{,k}-\tfrac14 E^j_{a'}e^{b'}_{j;k}\ga_{b'}\ga^{a'}\chi +  A_k\chi\}
\end{aligned}
\end{equation}
\br
The term in the braces is the covariant derivative of $\chi$ with respect to the spin connection $\tilde\Clk_k$
\begin{equation}\lae{3.44} 
\tilde\Clk^{b'}_{ka'}=\tfrac14\tilde\om^{b'}_{ka'}=-\tfrac14 E^j_{a'}e^{b'}_{j;k}\ga_{b'}\ga^{a'}
\end{equation}
and the Yang-Mills connection $(A_\mu)$ satisfying $A_0=0$ such that
\begin{equation}\lae{3.45}
\tilde D_k\chi=\chi_{,k}+\tilde\Clk_k\chi+ A_k\chi.
\end{equation}
The gauge transformations for both the Yang-Mills connection as well as for the spin connection do not depend on $x^0$ but only on $x\in\socc$. In case of the Yang-Mills connection this has already been proved in \cite[Lemma 2.6]{cg:uqtheory} while the proof for the spin connection $\tilde\Clk_k$ follows from \re{3.44} and \re{3.32} if we only consider Lorentzian metrics of the form
\begin{equation}
d\bar s^2=-dt^2+g_{ij}(x)dx^idx^j
\end{equation}
in a product manifold $N=I\times \socc$, as will be the case after the quantization of the Dirac field.
\er
Summarizing the preceding results we obtain:
\bl\lal{3.3}
The Dirac Lagrangian can be expressed in the form 
\begin{equation}\lae{3.46}
\begin{aligned}
L_D&=\tfrac{i}2(\bar\chi_I\dot\chi^I-\dot{\bar\chi}^{I}\chi_{I})w^{-1}\f^{-1}+m i\bar\chi_I\ga^0\chi^I\f^{-1}\\
&\q-\tfrac{i}2\{\bar\chi_I\ga^0E^k_{a'}\ga^{a'}\tilde D_k\chi^I-\overline{\bar\chi_I\ga^0E^k_{a'}\ga^{a'}\tilde D_k\chi^I}\}\f^{-1},
\end{aligned}
\end{equation}
where $\chi$ and $\tilde D_k$ are defined in \re{3.35} \resp \re{3.45}.  
\el

\section{Quantization of the Lagrangian}\las{4}
We consider the functional
\begin{equation}\lae{4.1}
\begin{aligned}
J&=\al_N^{-1}\int_{\tilde\Om}(\bar R-2\Lam)-\int_{\tilde\Om}\tfrac14 \ga_{\bar a\bar b}\bar g^{\mu\rho_2}\bar g^{\lam\rho_1}F^{\bar a}_{\mu\rho_1}F^{\bar b}_{\rho_2\lam}\\
&\hp =\;\;\,-\int_{\tilde\Om}\{\tfrac12 \bar g^{\mu\lam}\ga_{\bar a\bar b}\F^{\bar a}_\mu\bar\F^{\bar b}_\lam+V(\F)\}\\
&\hp =\;\;+\int_{\tilde\Om}\{\tfrac12[\tilde\psi_{I}E^\mu_a\ga^a(D_\mu \psi)^{I}+\overline{\tilde\psi_{I}E^\mu_a\ga^a(D_\mu \psi)^{I}}]+m\tilde\psi_{I}\psi^{I}\},
\end{aligned}
\end{equation}
where $\al_N$ is a positive coupling constant and $\tilde\Om\Su N$.  

We use the action principle that, for an arbitrary $\tilde\Om$ as above, a solution $(A,\F,\psi,\bar g)$ should be a stationary point of the functional with respect to compact variations. This principle requires no additional surface terms for the functional.

As we proved in \cite{cg:qgravity} we may only consider metrics $\bar g_{\al\bet}$ that split with respect to some fixed globally defined time function $x^0$ such that
\begin{equation}\lae{4.2}
d\bar s^2=-w^2 (dx^0)^2+g_{ij}dx^idx^j
\end{equation}
where $g(x^0,\cdot)$ are Riemannian metrics in $\socc$,
\begin{equation}
\socc=\{x^0=0\}.
\end{equation}
The first functional on the right-hand side of \re{4.1} can be written in the form  
\begin{equation}\lae{5.4}
\al^{-1}_N\int_a^b\int_\Om\{\tfrac14G^{ij,kl}\dot g_{ij}\dot g_{kl}w^{-2}+R-2\Lam\}w\f,
\end{equation}
where 
\begin{equation}
G^{ij,kl}=\tfrac12\{g^{ik}g^{jl}+g^{il}g^{jk}\}-g^{ij}g^{kl}
\end{equation}
is the DeWitt metric,
\begin{equation}
(g^{ij})=(g_{ij})^{-1},
\end{equation}
$R$ the scalar curvature of the slices 
\begin{equation}
\{x^0=t\}
\end{equation}
with respect to the metric $g_{ij}(t,\cdot)$, and where we also assumed that $\tilde\Om$ is a cylinder
\begin{equation}
\tilde\Om=(a,b)\times\Om,\qq\Om\Su \socc,
\end{equation}
such that $\tilde\Om\su U_k$ for some $k\in \N$, where the $U_k$ are  special coordinate patches of $N$ such that there exists a local trivialization in $U_k$ with the properties that there is a fixed  Yang-Mills connection 
\begin{equation}
\bar A=(\bar A^{\bar a}_\mu)=f_{\bar a}\bar A^{\bar a}_\mu dx^\mu
\end{equation}
satisfying
\begin{equation}
\bar A^{\bar a}_0=0\qq \text{in}\; U_k,
\end{equation}
\cf \cite[Lemma 2.5]{cg:uqtheory}. We may then  assume that the Yang-Mills connections $A=(A^{\bar a}_\mu)$ are of the form
\begin{equation}
A^{\bar a}_\mu (t,x)=\bar A^{\bar a}_\mu (0,x)+\tilde A^{\bar a}_\mu (t,x),
\end{equation}
where $(\tilde A^{\bar a}_\mu )$ is a tensor, see \cite[Section 2]{cg:uqtheory}.

The Riemannian metrics $g_{ij}(t,\cdot)$ are elements of the bundle $T^{0,2}(\socc)$. Denote by $E$ the fiber bundle with base $\socc$ where the fibers $F(x)$ consists of the Riemannian metrics $(g_{ij})$. We shall consider each fiber to be a Lorentzian manifold equipped with the DeWitt metric. Each fiber $F$ has dimension
\begin{equation}
\dim F=\frac{n(n+1)}2\equiv m+1.
\end{equation}
Let $(\xi^r)$, $0\le r\le m$, be  coordinates for a local trivialization such that
\begin{equation}
g_{ij}(x,\xi^r)
\end{equation}
is a local embedding. The DeWitt metric is then expressed as
\begin{equation}
G_{rs}=G^{ij,kl}g_{ij,r}g_{kl,s},
\end{equation}
where a comma indicates partial differentiation.  
In the new coordinate system the curves 
\begin{equation}
t\ra g_{ij}(t,x)
\end{equation}
can be written in the form
\begin{equation}
t\ra \xi^r(t,x)
\end{equation}
and we infer
\begin{equation}
G^{ij,kl}\dot g_{ij}\dot g_{kl}=G_{rs}\dot\xi^r\dot\xi^s. 
\end{equation}
Hence, we can express \re{5.4} as 
\begin{equation}\lae{3.49}
J=\int_a^b\int_\Om \al_n^{-1}\{\tfrac14 G_{rs}\dot\xi^r\dot\xi^sw^{-1}\f+(R-2\Lam)w\f\},
\end{equation}
where we now refrain from writing down the density $\sqrt\rho$ explicitly, since it does not depend on $(g_{ij})$ and therefore should not be part of the Legendre transformation.  Here we follow Mackey's advice in \cite[p. 94]{mackey:book} to always consider rectangular coordinates when applying canonical quantization, which can be rephrased that the Hamiltonian has to be a coordinate invariant, hence no densities are allowed.

Denoting the Lagrangian \tit{function} in \re{3.49} by $L$, we define
\begin{equation}
\pi_r= \pde L{\dot\xi^r}=\f G_{rs}\frac1{2\al_N}\dot\xi^sw^{-1}
\end{equation}
and we obtain for the Hamiltonian function $\hat H_G$
\begin{equation}\lae{5.18}
\begin{aligned}
\hat H_G&=\dot\xi^r\pde L{\dot\xi^r}-L\\
&=\f G_{rs}\big(\frac1{2\al_N}\dot\xi^rw^{-1}\big)\big(\frac1{2\al_N}\dot\xi^sw^{-1}\big) w\al_N-\al_N^{-1}(R-2\Lam)\f w\\
&=\f^{-1}G^{rs}\pi_r\pi_s w\al_N-\al^{-1}_N(R-2\Lam)\f w\\
&\equiv H_G w,
\end{aligned}
\end{equation}
where $G^{rs}$ is the inverse metric. Hence,
\begin{equation}
H_G=\al_N\f^{-1}G^{rs}\pi_r\pi_s-\al_N^{-1}(R-2\Lam)\f
\end{equation}
is the Hamiltonian that will enter the Hamilton constraint, for details see \cite[Chapter 1.4]{cg:qgravity-book}.

Let us recall that the fibers $F$ can be considered to be Lorentzian manifolds, even globally hyperbolic manifolds, equipped with the DeWitt metric $(\f G^{ij,kl})$, where $\f$ is a time function, \cf \cite[Theorem 1.4.2]{cg:qgravity-book}. In the fibers we can introduce new coordinates, $(\xi^a)=(\xi^0,\xi^A)\equiv (t,\xi^A)$ , $0\le a\le m$, and $1\le A\le m$, such that
\begin{equation}\lae{4.22}
t=\sqrt\f
\end{equation}
and $(\xi^A)$ are coordinates for the hypersurface
\begin{equation}
M=\{\f=1\}=\{\xi^0=1\}.
\end{equation}
The Lorentzian metric in the fibers can then be expressed in the form
\begin{equation}\lae{4.24}
ds^2=-\frac{16(n-1)}n dt^2+t^2 G_{AB}d\xi^Ad\xi^B,
\end{equation}
where $(G_{AB})$ is a Riemannian metric on $M$ which is independent of $t$. When we work in a local trivialization of the bundle $E$ the coordinates $(\xi^A)$ are independent of $x$. The time coordinate $t$ is also independent of $x$, \cf \cite[Lemma 1.8]{cg:qgravity}. Moreover, the fiber elements $(g_{ij})$ can be expressed in the form
\begin{equation}\lae{4.25}
g_{ij}=t^\frac4n \s_{ij},
\end{equation}
where $(\s_{ij})$ is an element of $M$, i.e., 
\begin{equation}
t(\s_{ij})=1,
\end{equation}
or equivalently,
\begin{equation}
\det{\s_{ij}}=\det{\rho_{ij}}.
\end{equation}

Next, let us look at the Yang-Mills Lagrangian which can be expressed as
\begin{equation}\lae{3.39}
L_{YM}=\tfrac12\ga_{\bar a\bar b}g^{ij}\tilde A^{\bar a}_{i,0}\tilde A^{\bar b}_{j,0}w^{-1}\f-\tfrac14 F_{ij}F^{ij}w\f.
\end{equation}
Let $E_0$ be the adjoint bundle
\begin{equation}
E_0=(S_0,\mf g,\pi,\Ad(\mc G))
\end{equation}
with base space $\socc$, where the gauge transformations only depend on the spatial variables $x=(x^i)$. Then the mappings $ t\ra\tilde A^{\bar a}_i(t,\cdot)$ can be looked at as curves in $T^{1,0}(E_0)\otimes T^{0,1}(\socc)$, where
the fibers of $T^{1,0}(E_0)\otimes T^{0.1}(\socc)$ are the tensor products
\begin{equation}
\mf g\otimes T^{0,1}_x(\socc),\qq x\in \socc,
\end{equation}
which are vector spaces equipped with metric
\begin{equation}
\ga_{\bar a\bar b}\otimes g^{ij}. 
\end{equation}
For our purposes it is more convenient to consider the fibers to be Riemannian manifolds endowed with the above metric. Let $(\zeta^p)$, $1\le p\le n_1n$, where $n_0=\dim\mf g$, be  local coordinates and
\begin{equation}
(\zeta^p)\ra \tilde A^{\bar a}_i(\zeta^p)\equiv \tilde A(\zeta)
\end{equation}
be a local embedding, then the metric has the coefficients
\begin{equation}\lae{4.33}
G_{pq}=\spd{\tilde A_p}{\tilde A_q}=\ga_{\bar a\bar b}g^{ij}\tilde A^{\bar a}_{i,p}\tilde A^{\bar b}_{j,q}.
\end{equation}
Hence, the Lagrangian $L_{YM}$ in \re{3.39} can be expressed in the form
\begin{equation}
L_{YM}=\tfrac12G_{pq}\dot\zeta^p\dot\zeta^qw^{-1}\f-\tfrac14F_{ij}F^{ij}w\f
\end{equation}
and we deduce
\begin{equation}
\tilde\pi_p=\pde{L_{YM}}{\dot\zeta^p}=G_{pq}\dot\zeta^qw^{-1}\f
\end{equation}
yielding the Hamilton function
\begin{equation}
\begin{aligned}
\hat H_{YM}&=\pi_p\dot\zeta^p-L_{YM}\\
&=\tfrac12 G_{pq}(\dot\zeta^pw^{-1}\f)(\dot\zeta^qw^{-1}\f)w\f^{-1}+\tfrac14F_{ij}F^{ij}w\f\\
&=\tfrac 12G^{pq}\tilde\pi_p\tilde\pi_qw\f^{-1}+\tfrac14F_{ij}F^{ij}w\f\\
&\equiv H_{YM}w.
\end{aligned}
\end{equation}
Thus, after introducing a normal Gaussian coordinate system such that $w=1$,  the Hamiltonian that will enter the Hamilton constraint equation is
\begin{equation}
H_{YM}=\tfrac 12 \f^{-1}{G}^{pq}\tilde\pi_p\tilde\pi_q+\tfrac14F_{ij}F^{ij}\f.
\end{equation}

Combining, now, \re{4.22}, \re{4.25} and \re{4.33}  we infer that the Yang-Mills Hamiltonian can be expressed as
\begin{equation}\lae{3.51}
H_{YM}=t^{\frac4 n-2}\tilde G^{pq}\tilde\pi_p\tilde\pi_q+\tfrac14F_{ij}F^{ij} t^{2-\frac8n},
\end{equation}
where the indices in the last term are raised with respect to the metric $\s_{ij}$, i.e.,
\begin{equation}
F^{ij}=\s^{ik}\s^{jl} F_{kl}.
\end{equation}
In case\ann{c} $n=3$ the exponents of $t$ in \re{3.51} are equal
\begin{equation}
\frac43-2=2-\frac83=-\frac23
\end{equation}
and we can write
\begin{equation}
\begin{aligned}
H_{YM}&=t^{-\frac23}\{\tilde G^{pq}\tilde\pi_p\tilde\pi_q+\tfrac14F_{ij}F^{ij}\} \\
&\equiv t^{-\frac23}\tilde H_{YM}.
\end{aligned}
\end{equation}
Moreover, if $(\s_{ij})$ as well as $(\rho_{ij})$ would be equal to the Euclidean metric $(\de_{ij})$, then the quantization of $\tilde H_{YM}$ would be achieved by known methods of QFT.

Hence, we shall try to express the Hamiltonians of the other physical forces like the Dirac and Higgs Hamiltonians, when evaluated for 
\begin{equation}
\s_{ij}=\rho_{ij}=\de_{ij}
\end{equation}
and in case $n=3$ in the form
\begin{equation}\lae{4.43}
H_D=t^{-\frac23}\tilde H_D
\end{equation}
\resp
\begin{equation}\lae{4.44}
H_{H}=t^{-\frac23}\tilde H_{H}
\end{equation}
such the quantization of the spatial Hamiltonian
\begin{equation}
\tilde H_{YM}+\tilde H_D+\tilde H_{H}
\end{equation}
would be well known, and in the end, all spatial Hamiltonians of the Standard Model could be incorporated.

Let us first consider the Dirac Hamiltonian. In the Dirac Lagrangian $L_D$, defined in equation \fre{3.46}, the volume density $\sqrt g$ is missing, i.e., in order to define the Hamiltonian we have to multiply the Lagrangian with $\sqrt g$ or, since we would like to work with functions instead of densities, we have to multiply the Lagrangian with $\f$.

In addition we shall also consider---at least locally---a normal Gaussian coordinate system such that $w=1$. Then, the final Dirac Lagrangian has the form
\begin{equation}\lae{4.46}
\begin{aligned}
L_D&=\tfrac{i}2(\bar\chi_I\dot\chi^I-\dot{\bar\chi}^{I}\chi_{I})+m i\bar\chi_I\ga^0\chi^I\\
&\q-\tfrac{i}2\{\bar\chi_I\ga^0E^k_{a'}\ga^{a'}\tilde D_k\chi^I-\overline{\bar\chi_I\ga^0E^k_{a'}\ga^{a'}\tilde D_k\chi^I}\},
\end{aligned}
\end{equation}

The spinorial variables  $\chi^I_A$ are anticommuting Grassmann variables. They  are elements of a  Grassmann algebra with involution, where the involution corresponds to the complex conjugation and will be denoted by a bar. 

The $\chi^{I}_A$ are complex variables and we define its real \resp imaginary parts as 
\begin{equation}
\xi^{I}_A=\tfrac1{\sqrt 2}(\chi^{I}_A+\bar\chi^{I}_A)
\end{equation}
\resp
\begin{equation}
\h^{I}_A=\tfrac1{\sqrt 2 i}(\chi^{I}_A-\bar\chi^{I}_A).
\end{equation}
Then,
\begin{equation}\lae{4.6.1}
\chi^{I}_A=\tfrac1{\sqrt 2}(\xi^{I}_A+i\h^{I}_A)
\end{equation}
and
\begin{equation}\lae{4.7.1}
\bar\chi^{I}_A=\tfrac1{\sqrt 2}(\xi^{I}_A-i\h^{I}_A).
\end{equation}

With these definitions we obtain
\begin{equation}
\frac{i}2(\bar\chi_{I}\dot\chi^{I}-\bar{\dot\chi}^{I}\chi_{I})=\frac{i}2(\xi^A_{I}\dot\xi^{I}_A+\h^A_{I}\dot\h^{I}_A).
\end{equation}

Casalbuoni  quantized a Bose-Fermi system in \cite[section 4]{casalbuoni:fermi} the results of which can be applied to spin $\frac12$ fermions. The Lagrangian in \cite{casalbuoni:fermi} is the same as the main part  our Lagrangian in \fre{4.46}, and the left derivative is used in that paper, hence we are using left derivatives as well such that  the conjugate momenta of the odd variables are, e.g.,
\begin{equation}
\pi^A_{I}=\frac{\pa L}{\pa \dot\xi^{I}_A}=-\frac i2\xi^A_{I},
\end{equation}
and thus the conclusions in \cite{casalbuoni:fermi} can be applied. 

The Lagrangian has been expressed in real variables---at least the important part of it---and it follows that the odd variables $\xi^{I}_A,\h^{I}_A$ satisfy, after introducing anticommutative Dirac brackets as in \cite[equ. (4.11)]{casalbuoni:fermi}, 
\begin{equation}
\{\xi_{I}^A,\xi^{J}_B\}^*_+=-i\de^{J}_I\de_{B}^A,
\end{equation}
\begin{equation}
\{\h_{I}^A,\h^{J}_B\}^*_+=-i\de^{J}_I\de_{B}^A,
\end{equation}
and
\begin{equation}
\{\xi_{I}^A,\h^{J}_B\}^*_+=0,
\end{equation}
\cf \cite[equ. (4.19)]{casalbuoni:fermi}.

In view of \re{4.6.1}, \re{4.7.1} we then derive
\begin{equation}
\{\bar\chi_{I}^A,\chi^{J}_B\}^*_+=-i\de^{J}_I\de_{B}^A,
\end{equation}
where $\bar\chi_{I}^A$ are the conjugate momenta.

Canonical quantization---with $\bar h=1$---then requires that the corresponding operators $\hat\chi^{I}_A, \hat{\bar\chi}_{J}^B$ satisfy the anticommutative rules 
\begin{equation}\lae{4.15.1}
[\hat\chi^{I}_A,\hat{\bar\chi}_{J}^B]_+=i\{\chi^{I}_A,\bar\chi_{J}^B\}^*_+=\de^{I}_J\de_{A}^B
\end{equation}
and
\begin{equation}
[\hat{\bar\chi}_{I}^A,\hat{\bar\chi}_{J}^B]_+=[\hat\chi^{I}_A,\hat{\chi}^{J}_B]_+=0,
\end{equation}
\cf \cite[equ. (3.10)]{casalbuoni:odd} and \cite[equ. (5.17)]{casalbuoni:fermi}.

From \re{4.46} we then deduce that the spinorial Hamilton function is equal to
\begin{equation}\lae{4.65} 
\begin{aligned}
 H_D&=
\tfrac{i}2\{\bar\chi_I\ga^0E^k_{a'}\ga^{a'}\tilde D_k\chi^I-\overline{\bar\chi_I\ga^0E^k_{a'}\ga^{a'}\tilde D_k\chi^I}\}\\
&\q-m i\bar\chi_I\ga^0\chi^I.
\end{aligned}
\end{equation}
When we try to quantize this Hamilton function then the vielbein $e^{a'}_k$ and its inverse $E^k_{a'}$ will correspond to a given element  $g_{ij}(x)$ in the fiber $F$ which can be expressed as in \re{4.25} and we deduce that the vielbein
\begin{equation}
\tilde e^{a'}_k=t^{-\frac2n}e^{a'}_k
\end{equation}
and its inverse
\begin{equation}
\tilde E^k_{a'}=t^{\frac2n}E^k_{a'}
\end{equation}
correspond to the metric $\s_{ij}$. Furthermore, the covariant derivative $\tilde D_k\chi^I$ is independent of $t$, in view of \re{3.44} and \fre{3.45}. Thus, the Hamilton function $H_D$ can expressed as
\begin{equation}\lae{4.65.1} 
\begin{aligned}
 H_D&=
t^{-\frac23}\big (\tfrac{i}2\{\bar\chi_I\ga^0\tilde E^k_{a'}\ga^{a'}\tilde D_k\chi^I-\overline{\bar\chi_I\ga^0\tilde E^k_{a'}\ga^{a'}\tilde D_k\chi^I}\}\big)\\
&\q-m i\bar\chi_I\ga^0\chi^I,
\end{aligned}
\end{equation}
i.e., the main  part has already the form that we looked for in \re{4.43}, provided $n=3$, only the mass term spoils the necessary configuration. To overcome this setback we either have to omit the mass term or to modify it by multiplying the mass term  in \fre{1.1} with the factor
\begin{equation}
\f^{-\frac1n},
\end{equation}
where $\f$ is defined in \fre{3.34.1}. Note that $\f=1$ if
\begin{equation}
g_{ij}=\rho_{ij}=\de_{ij}
\end{equation}
as is the case in QFT. Anyway, either by omitting or by modifying the mass term the Dirac Hamilton function can be expressed in the required form\ann{$t^{-\frac23}$}
\begin{equation}
H_D=t^{-\frac23}\tilde H_D,
\end{equation}
where the underlying Riemannian metric is $\s_{ij}$ provided $n=3$.

The remaining Hamiltonian is the Hamiltonian of the Higgs field. The Higgs Lagrangian is defined by
\begin{equation}\lae{4.3.1}
L_H=-\tfrac12\bar g^{\al\bet}\ga_{ab}\F^a_\al\F^b_\bet-V(\F),
\end{equation}
where $V$ is a smooth potential. We assume that in a local coordinate system $\F$ has real coefficients. The covariant derivatives of $\F$ are defined by a connection $A=(A^a_\mu)$ in $E_1$ 
\begin{equation}
\F^a_\mu=\F^a_{,\mu}+f^a_{cb}A^c_\mu\F^b.
\end{equation}
As in the preceding section we work in a local trivialization of $E_1$ using the temporal gauge, i.e.,
\begin{equation}
A^a_0=0,
\end{equation}
hence, we conclude
\begin{equation}
\F^a_0=\F^a_{,0}.
\end{equation}

Expressing the density $g$ as in \fre{3.34.1} we obtain Lagrangian 
\begin{equation}
L_H=\tfrac12 \ga_{ab}\F^a_{,0}\F^b_{,0}\f-\tfrac12g^{ij}\ga_{ab}\F^a_i\F^b_j\f-V(\F)\f,
\end{equation}
where, again, we used local coordinates such that $w=1$. In order to apply our approach, outlined in \re{4.44}, we have to modify the Lagrangian. Instead of the above Lagrangian we have to consider
\begin{equation}
L_{H mod}=\{\tfrac12 \ga_{ab}\F^a_{,0}\F^b_{,0}-\tfrac12g^{ij}\ga_{ab}\F^a_i\F^b_j\}\f^{1+\ga_1}-V(\F)\f^{1+\ga_2}.
\end{equation}
Let us define
\begin{equation}
p_a=\pde{L_H}{\dot\F^a},\qq\dot\F^a=\F^a_{,0},
\end{equation}
then we obtain the Hamilton function
\begin{equation}
\begin{aligned}
 H_{H mod}&=p_a\dot\F^a-L_H\\
&=\tfrac12\ga^{ab}p_ap_b\f^{-(1+\ga_1)}+\tfrac12 g^{ij}\ga_{ab}\F^a_i\F^b_j\f^{1+\ga_1}+V(\F)\f^{1+\ga_2}.
\end{aligned}
\end{equation}
After quantization the $g_{ij}$ are elements of the fiber $F$, i.e.,
\begin{equation}
g_{ij}=t^\frac4n\s_{ij}.
\end{equation}
If $n=3$, then $\ga_1$ has to be chosen such that
\begin{equation}
-2(1+\ga_1)=-\frac43+2(1+\ga_1)=-\frac23
\end{equation}
which is the case if
\begin{equation}
\ga_1=-\frac23.
\end{equation}
For $\ga_2$ we obtain
\begin{equation}
2(1+\ga_2)=-\frac23
\end{equation}
yielding
\begin{equation}
\ga_2=-\frac43.
\end{equation}

Thus, the Hamilton function of the modified Higgs field has the required form
\begin{equation}\lae{4.10}
H_{H mod}=t^{-\frac23}\tilde H_{H mod},
\end{equation}
where
\begin{equation}
\tilde H_{Hmod}=\tfrac12\ga^{ab}p_ap_b+\tfrac12 \s^{ij}\ga_{ab}\F^a_i\F^b_j+V(\F)
\end{equation}
is a standard Hamiltonian of a Higgs field in QFT by choosing $\s_{ij}=\de_{ij}$ and $\F$,  $V(\F)$ as well as the Yang-Mills connection appropriately.

Combining the four Hamilton functions in \re{5.18}, \re{3.51}, \re{4.10} and \re{4.65.1} the Hamilton constraint has the form
\begin{equation}\lae{4.81}
\begin{aligned}
H&=H_G+H_{YM}+H_H+H_D\\
&=H_G+t^{-\frac23}(\tilde H_{YM}+\tilde H_H+\tilde H_D)\\
&\equiv H_G+t^{-\frac23} \tilde H_{SM}=0,
\end{aligned}
\end{equation}
where we omitted the subscript $mod$ and where $SM$ refers to the fields  of the Standard Model or to a  corresponding subset of fields.

The Hamiltonian 
\begin{equation}
H_G=\al_N\f^{-1}G^{rs}\pi_r\pi_s-\al_N^{-1}(R-2\Lam)\f
\end{equation}
we quantize as in our former papers \cite{cg:qgravity} and \cite{cg:qgravity2b} to obtain
\begin{equation}\lae{4.83}
H_G=-\al_N \D-\al_N^{-1}R t^2+ 2\al_N^{-1}\Lam t^2,
\end{equation}
where the Laplacian is the Laplacian of the metric \re{4.24} acting\ann{action} in the fibers $F$ of $E$. The Laplacian acts on smooth functions $u$ of the form $u=u(g_{ij})$. Choosing the Gaussian  coordinate system $(\xi^a)=(t,\xi^A)$ such that the fiber metric has form as in \re{4.24}, then, the hyperbolic term $-\D u$ can be expressed as
\begin{equation}
-\D u=\frac n{16(n-1)}t^{-m}\pde{}{t}(t^m \pde ut)-t^{-2}\bar\D u,
\end{equation}
where $\bar\D$ is the Laplacian of the hypersurface 
\begin{equation}
M=\{t=1\}.
\end{equation}
Using separation of variables we consider functions $u$ which are products
\begin{equation}\lae{3.57}
u(t,\xi^A)=w(t)  v(\xi^A),
\end{equation}
where $v$ is a spatial eigenfunction, or eigendistribution, of the Laplacian $\bar\D$
\begin{equation}
-\bar\D v=\lam v.
\end{equation}

The hypersurface
\begin{equation}
M=\{\f=1\}
\end{equation}
can be considered to be a subbundle of $E$, where each fiber $M(x)$ is a hypersurface in the fiber $F(x)$ of $E$. We shall use the same notation $M$ for the  subbundle as well as for the hypersurface and in general we shall omit the reference to the base point $x\in \socc$. Furthermore, we specify the metric $\rho_{ij}\in T^{0,2}(\socc)$, which we used to define $\f$, to be equal to the Euclidean metric such that in Euclidean coordinates
\begin{equation}
\f^2=\frac{\det g_{ij}}{\det \de_{ij}}=\det g_{ij}.
\end{equation}
Then, it is well-known that each $M(x)$ with the induced metric $(G_{AB})$ is a symmetric space, namely, it is isometric to the coset space 
\begin{equation}\lae{4.3}
G/K=SL(n,\R[])/SO(n),
\end{equation}
\cf \cite[equ.(5.17), p. 1123]{dewitt:gravity} and \cite[p. 3]{jorgenson:book}. The eigenfunctions in symmetric spaces, and especially of the coset space in \re{4.3}, are well-known, they are the so-called \tit{spherical functions}. One can also define a Fourier transformation for functions in $L^2(G/K)$ and prove a Plancherel formula, similar to the Euclidean case, \cf \cite[Chapter III]{helgason:ga}. Also similar to the Euclidean case we shall use the Fourier kernel to define the eigenfunctions, or eigendistributions, \cf \cite[Section 5]{cg:qgravity3}. 

Let
\begin{equation}\lae{4.9}
G=NAK
\end{equation}
be an Iwasawa decomposition of $G$, where $N$ is the subgroup of unit upper triangle matrices, $A$ the abelian subgroup of diagonal matrices with strictly positive diagonal components and $K=SO(n)$. The corresponding Lie algebras are denoted by
\begin{equation}
\mf g, \mf n,\mf a\tup{\, and\, }\mf k.
\end{equation}
Here,
\begin{equation}
\begin{aligned}
\mf g&= \tup{real matrices with zero trace}\\
\mf n&=\tup{subspace of strictly upper triangle matrices with zero diagonal}\\
\mf a&=\tup{subspace of diagonal matrices with zero trace}\\
\mf k&=\tup{subspace of skew-symmetric matrices}.
\end{aligned}
\end{equation}
The Iwasawa decomposition is unique. When
\begin{equation}
g=nak
\end{equation}
we define the maps $n, A, k$ by
\begin{equation}\lae{4.13}
g=n(g)A(g)k(g).
\end{equation}
We also use the expression $\log A(g)$, where $\log$ is the matrix logarithm. In case of diagonal matrices
\begin{equation}
a=\diag(a_1,\dots, a_n)
\end{equation}
with positive entries
\begin{equation}
\log a=\diag(\log a_i),
\end{equation}
hence
\begin{equation}
A(g)=e^{\log A(g)}.
\end{equation}
\br\lar{4.1}
(i) The Lie algebra $\mf a$ is a (n-1)-dimensional real algebra, which, as a vector space, is equipped with a natural real, symmetric scalar product, namely, the trace form
\begin{equation}
\spd{H_1}{H_2}=\tr (H_1H_2),\qq H_i\in\mf a.
\end{equation}

\cvm
(ii) Let $\mf a^*$ be the dual space of $\mf a$. Its elements will be denoted by Greek symbols, some of which have a special meaning in the literature. The linear forms are also called \tit{additive characters}.

\cvm
(iii) Let $\lam\in \mf a^*$, then there exists a unique matrix $H_\lam\in\mf a$ such that
\begin{equation}
\lam(H)=\spd{H_\lam}H\qq\A\, H\in \mf a.
\end{equation}
This definition allows to define a dual trace form in $\mf a^*$ by setting for $\lam,\mu\in \mf a^*$
\begin{equation}\lae{4.25.1}
\spd \lam \mu=\spd{H_\lam}{H_\mu}.
\end{equation}

\cvm
(iv) The Lie algebra $\mf g$ is a direct sum
\begin{equation}
\mf g=\mf n+\mf a+\mf k.
\end{equation}
Let $E_{ij}$, $1\le i<j\le n$, be the matrices with component $1$ in the entry $(i,j)$ and other components zero, then these matrices form a basis of $\mf n$. For $H\in \mf a$, $H=\diag (x_i)$, the Lie bracket in $\mf g$, which is simply the commutator, applied to $H$ and $E_{ij}$ yields
\begin{equation}
[H,E_{ij}]=(x_i-x_j)E_{ij}\q\A\, H\in\mf a.
\end{equation}
Hence, the $E_{ij}$ are the eigenvectors for the characters $\al_{ij}\in \mf a^*$ defined by
\begin{equation}\lae{4.104}
\al_{ij}(H)=x_i-x_j.
\end{equation}
Here, $E_{ij}$ is said to be an eigenvector of $\al_{ij}$, if 
\begin{equation}
[H,E_{ij}]=\al_{ij}(H) E_{ij}\qq\A\,H\in \mf a.
\end{equation}
The eigenspace of $\al_{ij}$ is one-dimensional. The characters $\al_{ij}$ are called the \tit{relevant} characters, or the $(\mf a,\mf n)$ characters. They are also called the positive restricted roots.
\er

The Fourier theory in $X=G/K$ which we have summarized in  \cite[Section 6]{cg:qgravity3} uses the functions
\begin{equation}\lae{4.102}
e_{\lam,b}(x)=e^{(i\lam+\rho)\log A(x,b)},\qq(\lam,b)\in \mf a^*\times B,\; x\in X,
\end{equation}
as the Fourier kernel, where
\begin{equation}
B=K/M.
\end{equation}
 Here, $M$ is the centralizer of $A$ in $K$ and $\rho$ a special character with norm
\begin{equation}
\spd\rho\rho=\frac1{12}(n-1)^2n,
\end{equation}
\cf \cite[Lemma 1]{cg:qgravity3}. If $n=3$ then
\begin{equation}\lae{4.109.1}
\abs\rho^2=1.
\end{equation}
For a precise definition of $A(x,b)\in A$ we refer to \cite[p.19]{cg:qgravity3}, where also references to the corresponding mathematical literature are given, especially to Helgason's book \cite[Chapter III]{helgason:ga}.

The Fourier transform for functions $f\in C^\un_c(X,\Cc)$ is then defined by
\begin{equation}\lae{4.62}
\hat f(\lam,b)=\int_Xf(x)e^{(-i\lam +\rho)\log A(x,b)}dx
\end{equation}
for $\lam\in\mf a^*$ and $b\in B$, or, if we use the definition in \re{4.102} 
\begin{equation}
e_{\lam,b}(x)=e^{(i\lam+\rho)\log A(x,b)},
\end{equation}
by
\begin{equation}
\hat f(\lam,b)=\int_Xf(x)\ol e_{\lam,b}(x)dx.
\end{equation}
The functions $e_{\lam,b}$ are real analytic in $x$ and are eigenfunctions of the Laplacian, \cf \cite[Prop. 3.14, p. 99]{helgason:ga},
\begin{equation}\lae{4.109}
-\tilde\D e_{\lam,b}=(\abs \lam^2+\abs\rho^2)e_{\lam,b},
\end{equation}
where 
\begin{equation}
\abs \lam^2=\spd \lam\lam,
\end{equation}
 \cf \re{4.25.1}, and similarly for $\abs\rho^2$. We also denote the Fourier transform by $\mc F$ such that
 \begin{equation}
\mc F(f)=\hat f.
\end{equation}
Its inverse $\mc F^{-1}$ is defined in $R(\mc F)$ by
\begin{equation}
f(x)=\frac 1{\abs W}\int_B\int_{\mf a^*}\hat f(\lam,b)\abs{\mf c(\lam)}^{-2}d\lam db,
\end{equation}
where $\mf c(\lam)$ is Harish-Chandra's  $\mf c$-function, $W$ is the Weyl group and
\begin{equation}
\abs W=\card W,
\end{equation}
the number of elements in $W$, in our case $\abs W=n!$.

In equation \re{4.109} we identified 
\begin{equation}
\tilde\D=\D_M=\D_X.
\end{equation}
In \cite{cg:qgravity3} we finally dropped  the embellishment and simply wrote $\D$ when referring to the above Laplacian but at the moment we refrain from doing so in order to avoid confusion.

We shall consider the eigenfunctions $e_{\lam,b}$ as tempered distributions of the Schwartz space $\msc S(X)$ and shall use their Fourier transforms
\begin{equation}
\hat e_{\lam,b}=\de_{(\lam,b)}=\de_\lam \otimes \de_b
\end{equation}
as the spatial eigenfunctions of
\begin{equation}
\mc F(-\D)=m(\mu)=(\abs \mu^2+\abs\rho^2),
\end{equation}
which is a multiplication operator, such that
\begin{equation}
\mc F(-\D)\hat e_{\lam,b}=m(\mu)\hat e_{\lam,b}=(\abs \lam^2+\abs\rho^2)\hat e_{\lam,b},
\end{equation}
\cf \cite[Section 6]{cg:qgravity3} for details.

Looking at the Fourier transformed eigenfunctions
\begin{equation}
\hat e _{\lam,b}=\de_\lam\otimes \de_b
\end{equation}
it is obvious that the dependence on $b$ has to be eliminated, since there is neither a physical nor a mathematical motivation to distinguish between $e_{\lam,b}$ and $e_{\lam,b'}$. We discarded the integration over $B$ in \cite[Section 6]{cg:qgravity3} and picked instead a special element $b_0\in B$, namely,
\begin{equation}\lae{6.63}
b_0=eM,\qq e=\id\in K,
\end{equation}
and only consider the eigenfunctions $e_{\lam,b_0}$ with corresponding Fourier transforms
\begin{equation}
\de_{\lam}\equiv \de_\lam\otimes \de_{b_0}=\hat e_{\lam,b_0}, \qq \lam\in \mf a^*.
\end{equation}
For a justification, see \cite[Lemma 4]{cg:qgravity3} and the arguments preceding the referenced Lemma.

The eigenfunctions $e_{\lam,b_0}$ depend on the characters $\lam\in \mf a^*$, but not all characters are physically relevant. For a definition of the physically relevant characters let me cite \cite[Remark 2, p. 18]{cg:qgravity3}:
\br\lar{4.3}
The  characters $\al_{ij}$, $1\le i<j\le n$,  in \re{4.102} will represent the \tit{elementary gravitons} stemming from the degrees of freedom in choosing the coordinates 
\begin{equation}
g_{ij},\qq 1\le i<j\le n,
\end{equation} 
of a metric tensor. The diagonal elements offer in general additional $n$ degrees of freedom, but in our case, where we  consider metrics satisfying
\begin{equation}
\det g_{ij}=1,
\end{equation}
only $(n-1)$  diagonal components can be freely chosen, and we shall choose the first $(n-1)$  entries, namely,
\begin{equation}
g_{ii},\qq1\le i\le n-1.
\end{equation}
The corresponding additive characters are named $\al_i, 1\le i\le n-1$, and are defined by
\begin{equation}
\al_i(H)=h_i,
\end{equation}
if
\begin{equation}
H=\diag(h_1,\ldots, h_n).
\end{equation}
The characters $\al_i$, $1\le i\le n-1$, and $\al_{ij}$ $1\le i<j\le n$, will represent the $\frac{(n+2)(n-1)}2$ \tit{elementary gravitons} at the character level. We shall normalize the characters by defining
\begin{equation}\lae{4.46.1}
\tilde\al_i=\norm{H_{\al_i}}^{-1}\al_i
\end{equation}
and
\begin{equation}\lae{4.47}
\tilde\al_{ij}=\norm{H_{\al_{ij}}}^{-1}\al_{ij}
\end{equation}
such that the normalized characters have unit norm, \cf \re{4.25.1}.
\er
We can now define the corresponding forms in $\mf a^*$ with arbitrary energy levels:
\bd\lad{5.9}
Let $\lam\in \R[]_+$ be arbitrary. Then we consider the characters
\begin{equation}
\lam \tilde{\al}_i\q \wed \q \lam \tilde{\al}_{ij},
\end{equation}
where we recall that the terms embellished by a tilde refer to the corresponding unit vectors. Then the eigenfunctions representing the elementary gravitons are $e_{\lam\tilde \al_i,b_0}$ and $e_{\lam\tilde\al_{ij},b_0}$.

The corresponding eigenvalue with respect to $-\tilde\D$ is $\abs\lam^2+\abs\rho^2$, where by a slight abuse of notation $\abs \lam^2 = \lam^2$ and $\abs \rho^2=\spd\rho\rho$. Note that $\abs\rho^2=1$ if $n=3$, \cf \re{4.109.1}.  

We would also like to define a \tit{zero-point energy} eigenfunction by choosing $\lam\in \mf a^*=0$. The corresponding eigenfunction would be $e_{0,b_0}$ satisfying
\begin{equation}
-\tilde\D e_{0,b_0}=\abs \rho^2 e_{0,b_0}=e_{0,b_0}.
\end{equation}
if $n=3$. 
\ed

We are now able to quantize the Hamiltonian $H$ in \re{4.81}. For brevity we denote the quantized Hamiltonians, which are operators, by using  the same symbols as for the Hamilton functions. The Hamilton operator $H_G$ we express as in \re{4.83}
\begin{equation}
\begin{aligned}
H_G u&=\ann{-}\al_N \frac n{16(n-1)}t^{-m}\pde{}{t}(t^m \pde wt)v-\al_N t^{-2}w\bar\D v\\
&\q -\al_N^{-1}t^{2-\frac4n}R(\s_{ij})wv + 2\al_N^{-1}\Lam t^2 wv,
\end{aligned}
\end{equation}
where we used the separation of variables in \re{3.57}, the form of the metric in \re{4.25}, namely,
\begin{equation}
g_{ij}=t^{\frac4n} \s_{ij}
\end{equation}
and the relation  between the  scalar curvatures of conformal metrics
\begin{equation}
R(g)=t^{-\frac4n} R(\s).
\end{equation}.

Let us recall that for the quantization of $\tilde H_{SM}$ we shall specify $\s_{ij}=\de_{ij}$, such that the spatial eigendistributions, or approximate eigendistributions, $\psi$ satisfying
\begin{equation}
\tilde H_{SM}\psi=\lam_1\psi,\qq \lam_1\ge 0
\end{equation}
can be derived by applying standard methods of QFT. We then solve the Wheeler-DeWitt equation
\begin{equation}
Hu=0
\end{equation}
not for all $(t,\s_{ij})\in \R[+]\times M$ but only for $(t,\de_{ij})$, where $t>0$ is arbitrary. Thus, we shall solve
\begin{equation}
-\tilde \D v=(\abs \lam^2+\abs\rho^2)v
\end{equation}
by using
\begin{equation}
v=e_{\lam,b_0}
\end{equation}
for arbitrary $\s_{ij}\in M$, but we shall evaluate $e_{\lam, b_0}$ only at $\s_{ij}=\de_{ij}$. Furthermore, we observe that for $x=gK\in X$ and $b=kM\in B$ we have
\begin{equation}
A(x,b)=A(gK,kM)=A(k^{-1}g),
\end{equation}
\cf \cite[equ. (202), p. 18]{cg:qgravity3},  hence, if $b=b_0$, i.e., if $k=e=\id$ then
\begin{equation}
A(x,b_0)=A(g).
\end{equation}
Moreover, let
\begin{equation}
\pi: G/K \ra M
\end{equation}
be the isometry, then
\begin{equation}
\pi(gK)=gg^*,
\end{equation}
where $g^*$ is the adjoint. Thus, if $g=(\de_{ij})=e$ we infer
\begin{equation}
\s_{ij}=\de_{ij}\in M\im  e_{\lam,b_0}(\s_{ij})=1,
\end{equation}
and we have proved:
\bt\lat{4.4}
Let $n=3$, $v=e_{\lam,b_0}$ and let $\psi$ be an  eigendistribution of $\tilde H_{SM}$ when $\s_{ij}=\de_{ij}$ such that
\begin{equation}
-\tilde \D e_{\lam,b_0}=(\abs\lam^2+1)e_{\lam,b_0},
\end{equation}
\begin{equation}
\tilde H_{SM}\psi=\lam_1\psi,\qq\lam_1\ge 0,
\end{equation}
and let $w$ be a solution of the  ODE  
\begin{equation}\lae{4.147}
\begin{aligned}
t^{-m}\pde{}{t}(t^m \pde wt)&+\frac{32}3 (\abs\lam^2+1)t^{-2}w+\frac{32}3\al_N^{-1}\lam_1 t^{-\frac23}w\\
& \q+\frac{64}3\al_N^{-2}\Lam t^2w=0
\end{aligned}
\end{equation}
then
\begin{equation}
u=w e_{\lam,b_0}\psi
\end{equation}
is a solution of the Wheeler-DeWitt equation
\begin{equation}
Hu=0,
\end{equation}
where $e_{\lam,b_0}$ is evaluated at $\s_{ij}=\de_{ij}$ and where we note that $m=5$.
\et
We shall refer to $e_{\lam,b_0}$ and $\psi$ as the spatial eigenfunctions and to $w$ as the temporal eigenfunction.

\br
We could also apply the respective Fourier transforms to $-\tilde \D e_{\lam,b_0}$ \resp $\tilde H_{SM}\psi$ and consider
\begin{equation}
w\hat e_{\lam,b_0}\hat\psi
\end{equation}
as the solution in Fourier space, where $\hat\psi$ would be expressed with the help of the ladder operators.
\er
In the next section we shall analyze the temporal eigenfunctions.

\section{Temporal eigenfunctions}\las{5}

The temporal eigenfunctions have to satisfy the ODE \re{4.147} or equivalently
\begin{equation}\lae{5.1}
\begin{aligned}
\Ddot w+5 t^{-1} \dot w&+\frac{32}3 (\abs\lam^2+1)t^{-2}w+\frac{32}3\al_N^{-1}\lam_1 t^{-\frac23}w\\
& \q+\frac{64}3\al_N^{-2}\Lam t^2w=0,
\end{aligned}
\end{equation}
where we used that $m=5$, since we assume $n=3$. Let us denote the other constants in front of the three lower order terms by $m_1$,\ann{i} $m_2^2$ \resp $m_3$, then the ODE looks like 
\begin{equation}\lae{5.2}
\Ddot w+5 t^{-1} \dot w+m_1 t^{-2}w+m_2^2 t^{-\frac23}w+m_3 t^2w=0,
\end{equation}
where
\begin{equation}\lae{5.3}
m_1\ge\frac{32}3,\q m_2\ge 0,\q m_3\in\R[].
\end{equation}
The ODE \re{5.2} has two linearly independent solutions which are smooth and defined for all $t>0$. However, if $m_2$ as well as $m_3$ are both different from zero, then the solution cannot be expressed by known functions like variants of the Bessel functions. Only if this is not valid the solutions can be expressed by known functions.
\bt\lat{5.1}
Assume $m_3=0$ and $m_2>0$, then the solutions of the ODE \re{5.2} are generated by
\begin{equation}
J(\tfrac32\sqrt{m_1-4}\,i,\tfrac32 m_2 t^\frac23) t^{-2}
\end{equation}
and
\begin{equation}
J(-\tfrac32\sqrt{m_1-4}\,i,\tfrac32 m_2 t^\frac23) t^{-2},
\end{equation}
where $J(\lam,t)$ is the Bessel function of the first kind.
\et
\bp
We used Mathematica to obtain these solutions. The verification that these functions are indeed solutions is straightforward.\ann{straight forward}
\ep
\bl\lal{5.2}
The solutions in the theorem above diverge to complex infinity if $t$ tends to zero and they converge to zero if $t$ tends to infinity.
\el
\bp
The results can be derived by looking at a series expansion of the corresponding Bessel functions near the origin \resp near infinity.
\ep
Next, let us consider the solutions when $m_2=0$ and $m_3\not=0$. Then we distinguish two cases $m_3>0$ \resp $m_3<0$. For a better distinction we shall express $m_3$ in the form
\begin{equation}
m_3=m_4^2,\q m_4>0,
\end{equation}
in the first case and as
\begin{equation}
m_3=-m_4^2, \q m_4>0,
\end{equation}
in the second case.
\bt\lat{5.3}
Assume $m_2=0$ and $m_3>0$, then the solutions of the ODE \re{5.2} are generated by the functions
\begin{equation}
J(\tfrac12\sqrt{m_1-4}\,i,\tfrac12 m_4 t^2) t^{-2}
\end{equation}
and
\begin{equation}
J(-\tfrac12\sqrt{m_1-4}\,i,\tfrac12 m_4 t^2) t^{-2},
\end{equation}
where $J(\lam,t)$ is the Bessel function of the first kind.
\et
Similarly we obtain in the second case:
\bt\lat{5.4}
Assume $m_2=0$ and $m_3<0$, then the solutions of the ODE \re{5.2} are generated by the functions
\begin{equation}
I(\tfrac12\sqrt{m_1-4}\,i,\tfrac12 m_4 t^2) t^{-2}
\end{equation}
and
\begin{equation}
I(-\tfrac12\sqrt{m_1-4}\,i,\tfrac12 m_4 t^2) t^{-2},
\end{equation}
where $I(\lam,t)$ is the modified Bessel function of the first kind. In Mathematica this function is denoted by \tup{BesselI[$\lam,t]$}.
\et
The arguments in the proof of \rt{5.1} also apply in case of \rt{5.3} and \rt{5.4}.
\bl\lal{5.5}
The  solutions in \rt{5.3} \resp \rt{5.4} diverge to complex infinity if $t$ tends to zero as well as if $t$ tends to infinity.
\el
\bp
Same arguments as in the proof of \rl{5.2} apply. 
\ep

\section{Conclusions}\las{6}

The temporal eigenfunctions in the theorems of the previous section all become unbounded if $t\ra 0$, which can be described as a big bang on a quantum level. Furthermore, if we consider $t<0$, then the functions
\begin{equation}
\tilde w(t)=w(-t),\qq t<0,
\end{equation}
also satisfy the ODE \re{5.1} for $t<0$, if we replace $t^{-\frac23}$ by $\abs t^{-\frac23}$,  i.e., they are also temporal eigenfunctions  if the light cone in $E$ is flipped.

Thus, we conclude
\bt\lat{6.1}
The quantum model we derived for gravity combined with the forces of the Standard Model can be described by products of spatial and temporal eigenfunctions of corresponding self-adjoint operators with a continuous spectrum. 

We have a zero-point energy state as a spatial eigendistribution of the gravitational Hamiltonian with smallest eigenvalue $\abs\rho^2=1$ which could be considered to be the source of the dark energy.

Furthermore, we have a big bang singularity in $t=0$. Since the same quantum model is also valid by switching from $t>0$ to $t<0$, with appropriate changes to the temporal eigenfunctions, one could argue that at the big bang two universes with different time orientations could have been created such that, in view of the CPT theorem, one was filled with matter and the other with anti-matter.
\et

\br
One of the reviewers raised two questions. First, he wondered about the logic to combine a low energy event, the quantization of  the fields of the Standard Model with a flat metric, with an high energy event, the quantization of gravity. As we have already pointed out in the introduction a unified quantization of gravity and matter fields leads to a hyperbolic equation of second order in a fiber space, where the main part of the hyperbolic operator acts in the fibers. The zero order terms of the operator contains the contributions of the quantized matter Hamiltonian and the interaction of gravity with matter fields occurs with the help of the fiber variables $(t,\s_{ij})$. The metric $\s_{ij}$ is used in the quantization of the matter fields. Looking at the spatial eigenfunction $v$ of the gravitational Hamiltonian and its eigenvalue, which expresses the energy, then, the eigenvalue is independent of the metric $\s_{ij}$ at which $v$ is evaluated and only the evaluation point  is relevant for the interaction, i.e., even if a non-flat metric $\s_{ij}$ would have been used in the quantization of the matter fields the contribution to the unified operator would not have changed qualitatively. Furthermore, as we have already mentioned in the introduction, due to the scalar curvature term $R$ we cannot expect to solve the Wheeler-DeWitt equation for all $(t,\s_{ij})$ if we use separation of variables, instead, we have to choose metrics with constant scalar curvature. Thus, we opted for $\s_{ij}=\de_{ij}$, also out of necessity because we could not quantize the matter field in a curved spacetime.

The second interaction with respect to the variable $t$, the quantum time, is realized in the ODE, where the contributions by the spatial gravitational \resp matter eigenfunctions and also by  the cosmological constant $\Lam$ have a power of $t$ as a multiplicative factor with different exponents. For small $t$ the gravitational energy  dominates because of the factor $t^{-2}$, for larger $t$ the matter energy dominates because the factor $t^{-\frac23}$, and if $\Lam\not=0$, then the cosmological constant dominates for very large $t$ because of the factor $t^2$. This is also reflected in the results of \rl{5.2} and \frl{5.5}.

The second question raised concerned the QFT renormalizability in this unified setting.

The quantization of gravity takes place in the fibers of $E$ while the quantization of the matter fields takes place in the base space $\socc=\R[n]$ which we equipped with the Euclidean metric for this task. Hence, the usual renormalization techniques can be used to deal with infinities. The fibers are ignored in this process.
\er

\br
The Academic Editor of the journal also requested some observational predictions of the theory presented in this paper.

In \rt{6.1}  we already offered possible answers to two open questions, namely, the source of the dark energy and why is matter dominating anti-matter.

The big bang is only predicted by the singularity of the Friedmann model, a classical theory. In this paper the big bang is predicted on a quantum level which is a more appropriate level because the big bang is certainly a quantum event.

Powerful gravitational waves might be caused by quantum gravitational forces like the collision of two black holes. If this is the case then they should satisfy an ODE similar to that we analyzed in \rs{5}. The patterns produced by the wave detectors should be similar to the plots produced by the solutions of the ODE in \fre{5.1}, though the scalar curvature term does not appear in the ODE since $R(\de_{ij})=0$ and in the case of black holes $R$ would be constant but different from zero, i.e., the ODE should contain a term, probably positive, with the factor $t^{\frac23}$ and most likely no contribution by the Standard Model fields.
\er

\bibliographystyle{hamsplain}

\begin{thebibliography}{10}

\bibitem{adm:old}
R.~Arnowitt, S.~Deser, and C.~W. Misner, \emph{The dynamics of general
  relativity}, Gravitation: an introduction to current research (Louis Witten,
  ed.), John Wiley, New York, 1962, pp.~227--265.

\bibitem{casalbuoni:odd}
R.~Casalbuoni, \emph{On the quantization of systems with anticommuting
  variables}, Il Nuovo Cimento A (1971--1996) \textbf{33} (1976), no.~1,
  115--125,
  {\href{http://dx.doi.org/10.1007/BF02748689}{doi:10.1007/BF02748689}}.

\bibitem{casalbuoni:fermi}
\bysame, \emph{{The classical mechanics for Bose-Fermi systems}}, Il Nuovo
  Cimento A (1971--1996) \textbf{33} (1976), no.~3, 389--431,
  {\href{http://dx.doi.org/10.1007/BF02729860}{doi:10.1007/BF02729860}}.

\bibitem{dewitt:gravity}
Bryce~S. DeWitt, \emph{{Quantum Theory of Gravity. I. The Canonical Theory}},
  Phys. Rev. \textbf{160} (1967), 1113--1148,
  {\href{http://dx.doi.org/10.1103/PhysRev.160.1113}{doi:10.1103/PhysRev.160.1113}}.

\bibitem{eguchi:book}
Tohru Eguchi, Peter~B. Gilkey, and Andrew~J. Hanson, \emph{Gravitation, gauge
  theories and differential geometry}, Phys. Rept. \textbf{66} (1980), 213,
  {\href{http://dx.doi.org/10.1016/0370-1573(80)90130-1}{doi:10.1016/0370-1573(80)90130-1}}.

\bibitem{cg:qgravity}
Claus Gerhardt, \emph{{The quantization of gravity in globally hyperbolic
  spacetimes}}, Adv. Theor. Math. Phys. \textbf{17} (2013), no.~6, 1357--1391,
  {\href{http://arXiv.org/abs/1205.1427}{arXiv:1205.1427}},
  {\href{http://dx.doi.org/10.4310/ATMP.2013.v17.n6.a5}{doi:10.4310/ATMP.2013.v17.n6.a5}}.

\bibitem{cg:uqtheory}
\bysame, \emph{{A unified quantum theory I: gravity interacting with a
  Yang-Mills field}}, Adv. Theor. Math. Phys. \textbf{18} (2014), no.~5,
  1043--1062, {\href{http://arXiv.org/abs/1207.0491}{arXiv:1207.0491}},
  {\href{http://dx.doi.org/10.4310/ATMP.2014.v18.n5.a2}{doi:10.4310/ATMP.2014.v18.n5.a2}}.

\bibitem{cg:qbh}
\bysame, \emph{{The quantization of a black hole}},  (2016),
  {\href{http://arXiv.org/abs/1608.08209}{arXiv:1608.08209}}.

\bibitem{cg:qbh2}
\bysame, \emph{{The quantization of a Kerr-AdS black hole}}, Advances in
  Mathematical Physics \textbf{vol. 2018} (2018), Article ID 4328312, 10 pages,
  {\href{http://arXiv.org/abs/1708.04611}{arXiv:1708.04611}},
  {\href{http://dx.doi.org/10.1155/2018/4328312}{doi:10.1155/2018/4328312}}.

\bibitem{cg:qgravity-book}
\bysame, \emph{{The Quantization of Gravity}}, 1st ed., Fundamental Theories of
  Physics, vol. 194, Springer, Cham, 2018,
  {\href{http://dx.doi.org/10.1007/978-3-319-77371-1}{doi:10.1007/978-3-319-77371-1}}.

\bibitem{cg:qgravity2b}
\bysame, \emph{{The quantization of gravity}}, Adv. Theor. Math. Phys.
  \textbf{22} (2018), no.~3, 709--757,
  {\href{http://arXiv.org/abs/1501.01205}{arXiv:1501.01205}},
  {\href{http://dx.doi.org/10.4310/ATMP.2018.v22.n3.a4}{doi:10.4310/ATMP.2018.v22.n3.a4}}.

\bibitem{cg:qgravity3}
\bysame, \emph{{The quantization of gravity: Quantization of the Hamilton
  equations}}, Universe \textbf{7} (2021), no.~4, 91,
  {\href{http://dx.doi.org/10.3390/universe7040091}{doi:10.3390/universe7040091}}.

\bibitem{helgason:ga}
Sigurdur Helgason, \emph{Geometric analysis on symmetric spaces}, Mathematical
  surveys and monographs; 39, American Math. Soc., 1994,
  {\href{http://dx.doi.org/10.1090/surv/039}{doi:10.1090/surv/039}}.

\bibitem{jorgenson:book}
Jay Jorgenson and Serge Lang, \emph{{Spherical Inversion on {SLn}(R)}},
  Springer New York, 2001,
  {\href{http://dx.doi.org/10.1007/978-1-4684-9302-3}{doi:10.1007/978-1-4684-9302-3}}.

\bibitem{mackey:book}
George~W. Mackey, \emph{The mathematical foundations of quantum mechanics: {A}
  lecture-note volume}, W., A. Benjamin, Inc., New York-Amsterdam, 1963.

\end{thebibliography}
\providecommand{\bysame}{\leavevmode\hbox to3em{\hrulefill}\thinspace}
\providecommand{\href}[2]{#2}



\end{document}